\documentclass[english,tightenlines,eqsecnum,amsmath,amssymb,aps,prd,superscriptaddress,nofootinbib,showpacs,floats]{article}
\usepackage{epsfig}
\usepackage{babel}
\usepackage{authblk}
\usepackage{amsmath,amsfonts,amssymb,multicol}
\usepackage[usenames]{xcolor}
\definecolor{AHZ}{rgb}{0.0,0.0,0.9}
\definecolor{AHZ1}{rgb}{1,0.0,0.1}
\usepackage[colorlinks,linkcolor=AHZ,citecolor=AHZ1,bookmarks]{hyperref}
\long\def\/*#1*/{}
\usepackage{color}
{\definecolor{RED}{rgb}{1,0,0}
\definecolor{GREEN}{rgb}{0,1,0}
\definecolor{BLUE}{rgb}{0,0,1}
\usepackage{graphicx,caption,subcaption}
\usepackage{epsfig}
\usepackage{babel}
\usepackage{authblk}
\usepackage{amsmath,amsfonts,amssymb,multicol}
\long\def\/*#1*/{}
\usepackage{graphicx,caption,subcaption}
\usepackage{rotating}
\usepackage[usenames]{xcolor}
\usepackage{makecell}
\usepackage{tablefootnote}
\usepackage{threeparttable}
\usepackage{footnote}
\def\nn{\nonumber\\}
\newcommand{\f}[2]{\frac{#1}{#2}}

\begin{document}

\title{Bouncing cosmological solutions from $f({\sf R,T})$ gravity}
\author[1]{Hamid Shabani,\thanks{h.shabani@phys.usb.ac.ir}}
\author[2]{Amir Hadi Ziaie,\thanks{ah.ziaie@gmail.com}}

\affil[1]{Physics Department, Faculty of Sciences, University of Sistan and Baluchestan, Zahedan, Iran}
\affil[2]{Department of Physics, Kahnooj Branch, Islamic Azad University, Kerman, Iran}
\date{\today}
%
\maketitle
\begin{abstract}
\noindent
In this work we study classical bouncing solutions in the context of $f({\sf R},{\sf T})={\sf R}+h({\sf T})$ gravity in a flat {\sf FLRW} background using a perfect fluid as the only matter content. Our investigation is based on introducing an effective fluid through defining effective energy density and pressure; we call this reformulation as the \lq\lq{}{\it effective picture}\rq\rq{}. These definitions have been already introduced to study the energy conditions in $f({\sf R},{\sf T})$ gravity. We examine various models to which different effective equations of state, corresponding to different $h({\sf T})$ functions, can be attributed. It is also discussed that one can link between an assumed $f({\sf R},{\sf T})$ model in the effective picture and the theories with generalized equation of state ({\sf EoS}). We obtain cosmological scenarios exhibiting a nonsingular bounce before and after which the Universe lives within a de-Sitter phase. We then proceed to find general solutions for matter bounce and investigate their properties. We show that the properties of bouncing solution in the effective picture of $f({\sf R},{\sf T})$ gravity are as follows: for a specific form of the $f({\sf R,T})$ function, these solutions are without any future singularities. Moreover, stability analysis of the nonsingular solutions through matter density perturbations revealed that except two of the models, the parameters of scalar-type perturbations for the other ones have a slight transient fluctuation around the bounce point and damp to zero or a finite value at late times. Hence these bouncing solutions are stable against scalar-type perturbations. It is possible that all energy conditions be respected by the real perfect fluid, however, the null and the strong energy conditions can be violated by the effective fluid near the bounce event. These solutions always correspond to a maximum in the real matter energy density and a vanishing minimum in the effective density. The effective pressure varies between negative values and may show either a minimum or a maximum.
\end{abstract}
%
\section{Introduction}\label{sec:intro}
Today, the standard cosmological model ({\sf SCM}) or the Big-Bang cosmology has become the most acceptable model which encompasses our knowledge of the Universe as a whole.  For this reason it  is called also the \lq\lq{}concordance model\rq\rq{}~\cite{Ostriker95}. This model which allows one to track the cosmological evolution of the Universe very well, has matured over the last century, consolidating its theoretical foundations with increasingly accurate observations. We can numerate a number of the successes of the {\sf SCM} at the classic level. For example, it accounts for the expansion of the Universe (Hubble law), the black body nature of cosmic microwave background ({\sf CMB}) within the framework of the {\sf SCM} can be understood and the predictions of light-element abundances which were produced during the nucleosynthesis. It also provides a framework to study the cosmic structure formation~\cite{Coles02}. However, though the {\sf SCM} works very well in fitting many observations, it includes a number of deficiencies and weaknesses. For instance some problems which are rooted in cosmological relics such as magnetic monopoles \cite{relicmonopole}, gravitons~\cite{relicgraviton}, moduli \cite{relicmoludi} and baryon asymmetry~\cite{relicbaryon}. Despite the self-consistency and remarkable success of the {\sf SCM} in describing the evolution of the Universe back to only one hundredth of a second, a number of unanswered questions remain regarding the initial state of the Universe, such as flatness and horizon problems~\cite{flathorizonpro}. Moreover, there are some unresolved problems related to the origin and nature of dark matter ({\sf DM})~\cite{DMPRO}. Notwithstanding the excellent agreement with the observational data there still exists a number of challenging open problems associated with the late time evolution of the Universe, namely the nature of dark energy ({\sf DE}) and cosmological constant problem \cite{DEPR}. Though the inflation mechanism has been introduced to treat some of the mentioned issues such as, the horizon, flatness and magnetic monopole problems at early Universe~\cite{inflationmech}, the {\sf SCM} suffers from a more fundamental issue, i.e., the {\it initial cosmological  singularity} that the existence of which has been predicted by the pioneering works of Hawking, Penrose and Geroch in 1960s, known as the singularity theorems~\cite{hawpen} and their later extensions by Tipler in 1978~\cite{Tipsing} and by Borde, Vilenkin and Guth in the 1990s~\cite{Bordeetal1990} (see also \cite{GianlucaCalcagni2017} for a comprehensive study). According to these theorems, a cosmological singularity is unavoidable if spacetime dynamics is described by General Relativity ({\sf GR}) and if matter content of the Universe obeys certain energy conditions. A singular state is an extreme situation with infinite values of physical quantities, like temperature, energy density, and the spacetime curvature from which the Universe has started its evolution at a finite past. The existence of such an uncontrollable initial state is irritating, since \lq\lq{}\textit{a singularity can be naturally considered as a source of lawlessness}\rq\rq{}~\cite{Novello08}. A potential solution to the issue of cosmological singularity can be provided by \lq\lq{}{\it non-singular bouncing cosmologies}\rq\rq{}~\cite{bouncecos}. Beside a huge interest in the solutions that do not display singular behavior, there can be more motivations to seek for non-singular cosmological models. The first reason for removing the initial singularity is rooted in the initial value problem since a consistent gravitational theory requires a well-posed Cauchy problem~\cite{WELLSING}. However, owing to the fact that the gravitational field diverges at a spacetime singularity, we could not have a well formulated Cauchy problem as we cannot set the initial values at a singular spatial hypersurface given by $t=const.$ Another related issue is that the existence of a singularity is inconsistent with the entropy bound $S/E=(2\pi R)/c\hbar$, where $S$, $E$, $R$, $\hbar$ and $c$ being entropy, proper energy, the largest linear dimension, Planck's constant and the velocity of light, respectively~\cite{Novello08}.
\par
During the past decades, models which describe bouncing behavior have been designed and studied as an approach to resolve the problem of initial singularity. These models suggest that the Universe existed even before the Big-Bang and underwent an accelerated contraction phase towards reaching a non-vanishing minimum radius. The transition from a preceding cosmic contraction regime to the current accelerating expansion phase (as already predicted in {\sf SCM}) is the so called \lq\lq{}{\it Big Bounce}\rq\rq{}. From this perspective, the idea that the expansion phase is preceded by a contraction phase paves a new way towards modeling the early Universe and thus, may provide a suitable setting to obviate some of the problems of the {\sf SCM} without the need to an inflationary scenario. Although an acceptable model can be considered as the one being capable of explaining the issues that have been treated by inflationary mechanism, e.g., most inflationary scenarios can give the scale-invariant spectrum of the cosmological perturbations~\cite{Contreras17}, problems of the {\sf SCM} may find solutions in the contracting regime before the bounce occurs. The horizon problem, for example, is immediately resolved if the far separated regions of the present Universe were in causal connection during the previous contraction phase. Similarly, the homogeneity, flatness, and isotropy of the Universe may also be addressed by having a smoothing mechanism in the contraction phase, see e.g., \cite{REVBOUNCE1123} for more details. Moreover, though the fine-tuning is required to keep a stable contracting regime, the nonsingular bounce succeeds in sustaining a nearly scale-invariant power spectrum~\cite{Xuepospec}.
\par
For several years, great effort has been devoted to the study of  bouncing cosmologies within different frameworks. The resulted cosmological models could be obtained at a classical level or by quantum modifications. Most of the efforts in quantum gravity are devoted to reveal the nature of the initial singularity of the Universe and to better understand the origin of matter,
non-gravitational fields, and the very nature of the spacetime. Non-singular bouncing solutions generically appear in loop quantum cosmology ({\sf LQC})~\cite{lqcbounce}, where the variables and quantization techniques of loop quantum gravity are  employed to investigate the effects of quantum gravity in cosmological spacetimes~\cite{lqbounce1,lqbounce2,Ashtekar11}. The recent large amount of works done within the loop quantum cosmology ({\sf LQC}) show that when the curvature of spacetime reaches the Planck scale, the Big-Bang singularity is replaced by a quantum Big-Bounce with finite density and spacetime curvature~\cite{lqbounce3}. Another approach, based on the de Broglie-Bohm quantum theory,
utilizes the wave function of the Universe in order to determine a quantum trajectory of the Universe through a bounce~\cite{deroglie}. In the framework of {\sf LQC} the semi-classical Friedmann equations receive corrections as~\cite{Ashtekar11,freqlqc}
\begin{align}
&H^{2}_{{\sf LQC}}=\frac{8\pi G}{3}\rho\left(1-\frac{\rho}{\rho_{{\sf max}}}\right)\label{Int1},\\
&\dot{H}_{{\sf LQC}}=-4\pi G(\rho+p)\left(1-2\frac{\rho}{\rho_{{\sf max}}}\right)\label{Int2}.
\end{align}
where, $\rho_{{\sf max}}\approx0.41\rho_{{\sf Pl}}$ and $\rho_{{\sf Pl}}=c^5/\hbar G^2$ being the Planck density~\cite{Ashtekar11,vanrhopl}. We note that, the relative magnitude of $\rho$ and $\rho_{{\sf max}}$ enables one to distinguish classical and quantum regimes. By a short qualitative inspection of the above equations the general feature of the bouncing behavior will be revealed: initially, the Universe were in contracting phase at which the matter density and
curvature are very low compared to the Planck scale. As the Universe contracts more, the maximum density is reached so that the quantum evolution follows the classical trajectory at low densities and curvatures but undergoes a quantum
bounce at matter density $\rho=\rho_{{\sf max}}$, where we have $H_{{\sf LQC}}=0$ and also $\dot{H}_{{\sf LQC}}=4\pi G (\rho+p)$. The quantum regime then joins on to the classical trajectory that was contracting to the future. Therefore, the quantum gravity effects create a non-singular transition from contraction to expansion and thus the Big-Bang singularity is replaced by a quantum bounce. Furthermore, we see that for all matter fields which satisfy the weak energy condition ({\sf WEC}) we have $\dot{H}_{{\sf LQC}}>0$. These two results are accounted for the general conditions for the existence of a bouncing solution. Moreover, nonsingular bouncing scenarios have been also reported in nonlocal gravity model~\cite{nonlocalgravity} where effective Friedmann equation with a slight difference to equation (\ref{Int1}) has been proposed exhibiting a bouncing-accelerating behavior. See also~\cite{nonlocgravit} for probing the issue of singularity avoidance in nonlocal gravitational theories.
\par
Another type of theories are called non-singular \lq\lq{}{\it matter bounce}\rq\rq{} scenarios which is a cosmological model with an initial state of matter-dominated contraction and a non-singular bounce~\cite{mattbounce}. Such a model provides an alternative to inflationary cosmology for generating the observed spectrum of cosmological fluctuations~\cite{REVBOUNCE1123,Brandenberger12,WE13}. In these theories some matter fields are introduced in such a way that the {\sf WEC} is violated in order to make $\dot{H}>0$ at the bounce. From equation (\ref{Int2}), it is obvious that putting aside the correction term leads to negative values for the time derivative of the Hubble parameter for all fluids which respect {\sf WEC}. Therefore, in order to obtain a bouncing cosmology it is necessary to either go beyond the {\sf GR} framework, or else to introduce new forms of matter which violate the key energy conditions, i.e., the null energy condition ({\sf NEC}) and the {\sf WEC}. For a successful bounce, it can be shown that within the context of {\sf SCM} the {\sf NEC} and thus the {\sf WEC}, are violated for a period of time around the bouncing point. In the context of matter bounce scenarios, many studies have been performed using quintom matter~\cite{Cai08}, Lee-Wick matter~\cite{LEEWICK}, ghost condensate field~\cite{Lin11}, Galileon fields~\cite{Qiu11Gal} and phantom field~\cite{phantomfield}. Cosmological bouncing models have also been constructed via various approaches to modified gravity such as $f({\sf R})$ gravity~\cite{Barragan09,Oikonomou14,Odintsov14,Odintsov151,Oikonomou15}, teleparallel $f({\sf T})$ gravity~\cite{Cai11}, brane world models~\cite{Kehagias99}, Einstein-Cartan theory~\cite{ecbounce}, Horava-Lifshitz gravity~\cite{horlifb}, nonlocal gravity~\cite{nonlocb} and others~\cite{Odintsov152}. There are also other cosmological models such as Ekpyrotic model~\cite{Khoury01} and string cosmology~\cite{Gasperini93} which are alternatives to both inflation and matter bounce scenarios.
\par
In the present work we study the existence of the bouncing solutions in the context of $f({\sf R},{\sf T})$ gravity. These type of theories have been firstly introduced in~\cite{Harko11} and later, their different aspects have been carefully studied and analyzed in~ \cite{Alvarenga13,Shabani13,Harko14,Shabani14,Singh14,Sharif14,Alves16,Sun16,Zaregonbadi161,Zaregonbadi162,
Shabani171,Shabani172,Shabani173,Shabani174,Moraes17,modelwormprd,Deb18,Moraes18,Sahoo18}.
In the current work we use an effective approach that we have introduced previously in~\cite{Shabani18}. In this method one defines an \lq\lq{} effective fluid\rq\rq{} endowed with the effective energy density $\rho_{{\sf (eff)}}$ and effective pressure $p_{{\sf (eff)}}$ allowing thus, to reformulate the $f({\sf R},{\sf T})$ gravity field equations. In a class of the minimally coupled form, i.e., $f({\sf R},{\sf T})={\sf R}+h({\sf T})$, one usually presumes $h({\sf T})$ functions and solves the resulted field equations. Instead, using the effective fluid description we obtain the $h({\sf T})$ function which corresponds to an {\sf EoS} defined as $p_{(\sf{eff})}=\mathcal{Y}(\rho_{(\sf{eff})})$. We therefore observe that the effective fluid picture may at  least be imagined as a mathematical translation of gravitational interactions between the actual matter fields and the curvature to an overall behavior attributed to a mysterious fluid with $p_{({\sf eff})}({\sf T})$ and $\rho_{({\sf eff})}({\sf T})$. In the current article we discuss three different classes of models specified by three different effective {\sf EoS}s or equivalently three different $h({\sf T})$ functions. We shall see that in these cases we obtain
\begin{align}
&\rho_{({\sf eff})}({\sf T})=\beta_{1}{\sf T}+\beta_{2}{\sf T}^{\gamma}+\tilde{\rho}_{({\sf eff})}\label{Int3},\\
&p_{({\sf eff})}({\sf T})=\lambda_{1}{\sf T}+\lambda_{2}{\sf T}^{\gamma}+\tilde{p}_{({\sf eff})}\label{Int4},
\end{align}
where, $\beta_{i},~\lambda_{i},~\gamma,~\tilde{\rho}_{({\sf eff})}$ and $\tilde{p}_{({\sf eff})}$ are some constants. Eliminating the trace between the effective density and pressure leaves us with an {\sf EoS} as follows
\begin{align}
\frac{\lambda _1 \left(\rho -\tilde{\rho}\right)-\beta _1 \left(p-\tilde{p}\right)}{\beta _2 \lambda _1-\beta _1 \lambda _2}=\left[\frac{\lambda _2 \left(\rho -\tilde{\rho}\right)-\beta _2 \left(p-\tilde{p}\right)}{\beta _1 \lambda _2-\beta _2 \lambda _1}\right]^{\gamma }\label{Int5},
\end{align}
where the subscript \lq\lq{}{\sf eff} \rq\rq{} has been dropped. In view of relation (\ref{Int5}), we may conclude that in $f({\sf R},{\sf T})={\sf R}+h({\sf T})$ gravity, interactions of a perfect fluid with the spacetime curvature can be mapped effectively onto the behavior of an exotic fluid obeying equation of state (\ref{Int5}). It is quiet interesting that a reduced form of (\ref{Int5}) has been introduced in~\cite{Stefancic05} and further studied in~\cite{Nojiri05}. Such a complicated {\sf EoS} has been introduced to study the cosmological implication of a model with a mixture of two different fluids, i.e., effective quintessence and effective phantom. Additionally, one can find other related works investigating some \lq\lq{}exotic\rq\rq{} fluids which follow various subfamily of (\ref{Int5}). These theories may be called \lq\lq{}{\it modified equation of state}\rq\rq{} ({\sf MEoS}) models.
This branch of research presumes a mysterious fluid(s) specified by an unusual {\sf EoS} with the hope of dealing with some unanswered questions in the cosmological realm. For example some relevant works in the literature can be addressed as follows; in~\cite{Barrow90} the author has employed an {\sf EoS} of the form $p=-\rho+\gamma \rho^{\lambda}$ in order to obtain power-law and exponential inflationary solutions. The case with $\lambda=1/2$ has been analyzed in~\cite{Stefancic052,Nojiri052,Nojiri05} to focus on the future expansion of the Universe. Emergent Universe models have been studied in~\cite{Mukherjee06} by taking into account an exotic component with $p=A\rho-B \rho^{1/2}$ and in~\cite{Contreras16} with $A=-1$. Different cosmological aspects of {\sf DE} with more simple form of {\sf EoS}, i.e, $p_{{\sf DE}}=\alpha(\rho_{{\sf DE}}-\rho_{0})$ have been investigated in~\cite{Babichev05} and the study of cosmological bouncing solutions can be found in~\cite{Contreras17}.

Therefore, recasting the $f({\sf R},{\sf T})$ field equations into the \lq\lq{}{\it effective picture}\rq\rq{} may provide a bridge to the cosmological models supported by {\sf MEoS}. Via this connection the problem of an exotic fluid turns into the problem of a usual fluid with exotic gravitational interactions. However, contrary to the former, in the latter case we start with a predetermined Lagrangian, i.e., $f({\sf R},{\sf T})$ gravitational Lagrangian.
The importance of the effective picture becomes more clear when one considers the energy conditions in $f({\sf R},{\sf T})$ gravity. As discussed in~\cite{Sharif13}, in $f({\sf R},{\sf T})$ gravity the energy conditions would be obtained for effective pressure and effective energy density. Therefore, it is reasonable to define a fluid as a source with effective pressure and energy density. As we shall see, the bouncing solutions in $f({\sf R},{\sf T})$ gravity (using only one perfect fluid in a flat {\sf FLRW} background) in the framework of our effective fluid approach, exhibit nonsingular properties such that in a finite value of the bounce time $t_{{\sf b}}$, non of the cosmological quantities would diverge. More exactly, as $t\to t_{{\sf b}}$ we observe that the scale factor decreases to a minimum non-vanishing value, i.e., $a\to a_{{\sf b}}$,  $H\big|_{t\rightarrow t_{\sf b}}\!\!\!\rightarrow 0$, $\rho\big|_{t\rightarrow t_{\sf b}}\!\!\!\rightarrow\rho_{{\sf b}}$, $\rho_{{\sf (eff)}}\big|_{t\rightarrow t_{\sf b}}\!\!\!\rightarrow0$ and $p_{{\sf (eff)}}\big|_{t\rightarrow t_{\sf b}}\!\!\!\rightarrow{p_{{\sf(eff)}}}_{{\sf b}}$. Therefore, non of the future singularities would appear.
Also, in all cases we have $\mathcal{W}\big|_{t\rightarrow t_{\sf b}}\!\!\!\rightarrow-\infty$, where $\mathcal{W}$ being the effective {\sf EoS} parameter and subscript \lq\lq{}{\sf b}\rq\rq{} stands for the value of quantities at the time at which the bounce occurs. We then observe that if we want to describe nonsingular bouncing solutions in $f({\sf R},{\sf T})={\sf R}+h({\sf T})$ gravity using a minimally coupled scalar field, a phantom field should be employed. These solutions show a violation of the {\sf NEC} in addition to the strong energy condition {\sf SEC}. Such a behavior is predicted in {\sf GR} for a perfect fluid in {\sf FLRW} metric with $k=-1,0$~\cite{Molina99}.

The current research is planned as follows. In Sec.~\ref{Sec1} we briefly present the effective fluid picture. Sec.~\ref{Sec2} is devoted to the bouncing solutions with asymptotic de Sitter behavior before and after the bounce. We first analyze models with constant effective pressure in subsection~\ref{Sec21}, they are called models of type {\sf A}. We then proceed to investigate the corresponding bouncing solutions, the energy conditions, the scalar field representation and finally the stability of these type of  solutions. In subsection~\ref{Sec22} models which correspond to two different {\sf EoS}s assuming $p_{({\sf eff})}=\mathcal{Y}(\rho_{({\sf eff})})$ will be discussed. These models are named as {\sf B}, {\sf C}, {\sf D} and {\sf E} models. An example of the matter bounce solution is considered in Sec.~\ref{Sec4} which is labeled as model {\sf E}. The connection of  {\sf A}-{\sf E} models with {\sf MEoS} theories will be presented through the effective picture. Section~\ref{comm}, is devoted to study of scalar-type cosmological perturbations. In Sec. \ref{singularsols} we give a brief review of singular models in the context of $f({\sf R,T})$ gravity and obtain a class of solutions exhibiting singular behavior. Finally, in Sec.~\ref{con} we summarize our conclusions.

\section{Reformulation of $f({\sf R},{\sf T})$ field equations in terms of a conserved effective fluid}\label{Sec1}

In the present section we review the field equations of $f({\sf R},{\sf T})$ gravity theories and rewrite them in terms of a conserved effective fluid. This reformulation would allow us to better understand the properties of bouncing solutions as well as classifying them. In $f({\sf R},{\sf T})$ gravity, one usually chooses a mathematically suitable $f({\sf R},{\sf T})$ Lagrangian, then obtains the corresponding field equations, and finally tries to solve them. In ~\cite{Shabani18}, we introduced a novel point of view to dealing with the field equations of $f({\sf R},{\sf T})$ gravity. This approach is based on reconsidering the field equations in terms of a conserved effective fluid. One of the benefits of using this method is to obtain a form of $f({\sf R},{\sf T})$ function for a physically justified condition on the effective fluid. Therefore, instead of mathematical arbitrariness in selecting different $f({\sf R},{\sf T})$ functions, we have physically meaningful Lagrangian forms. Hence,  in this paper we make use of those $f({\sf R},{\sf T})$ functions that we obtained in our previous work~\cite{Shabani18}. The action integral in $f({\sf R},{\sf T})$ gravity theories is given by~\cite{Harko11}
\begin{align}\label{action}
{\sf S}=\int \sqrt{-{\sf g}} d^{4} x \left[\frac{1}{2\kappa^{2}} f\Big{(}{\sf R}, {\sf T}\Big{)}+{\sf L}^{{\sf (m)}}\right],
\end{align}
where ${\sf R}$, ${\sf T}\equiv g^{\mu \nu} {\sf T}_{\mu \nu}$, ${\sf L}^{\sf{\sf {(m)}}}$ denote the Ricci scalar,
the trace of energy momentum tensor ({\sf EMT}) and the Lagrangian of matter, respectively. The determinant of metric is shown by ${\sf g}$,
$\kappa^{2}\equiv8 \pi G$ is the gravitational coupling constant and we have set $c=1$. We assume that the matter Lagrangian ${\sf L}^{{\sf (m)}}$ depends only on the metric components therefore the following form for the {\sf EMT} can be defined
\begin{align}\label{Euler-Lagrange}
{\sf T}_{\mu \nu}\equiv-\frac{2}{\sqrt{-g}}
\frac{\delta\left[\sqrt{-{\sf g}}{\sf L}^{{\sf (m)}}\right]}{\delta {\sf g}^{\mu \nu}}.
\end{align}
By varying action (\ref{action}) with respect to the metric components ${\sf g}^{\alpha \beta}$ we obtain the following field equation~\cite{Harko11}
\begin{align}\label{fRT field equations}
&&F({\sf R},{\sf T}) {\sf R}_{\mu \nu}-\frac{1}{2} f({\sf R},{\sf T}) g_{\mu \nu}+\Big{(} g_{\mu \nu}
\square -\triangledown_{\mu} \triangledown_{\nu}\Big{)}F({\sf R},{\sf T})\nn&&=
\Big{(}\kappa^{2}-{\mathcal F}({\sf R},{\sf T})\Big{)}{\sf T}_{\mu \nu}-\mathcal {F}({\sf R},{\sf T})\mathbf
{\Theta_{\mu \nu}},
\end{align}
where we have defined
\begin{align}\label{theta}
\mathbf{\Theta_{\mu \nu}}\equiv g^{\alpha \beta}\frac{\delta
{\sf T}_{\alpha \beta}}{\delta {\sf g}^{\mu \nu}}=-2{\sf T}_{\alpha \beta}+{\sf g}_{\alpha \beta}{\sf L}^{{\sf (m)}},
\end{align}
and
\begin{align}\label{f definitions1}
{\mathcal F}({\sf R},{\sf T}) \equiv \frac{\partial f({\sf R},{\sf T})}{\partial {\sf T}}~~~~~
~~~~~\mbox{and}~~~~~~~~~~
F({\sf R},{\sf T}) \equiv \frac{\partial f({\sf R},{\sf T})}{\partial {\sf R}}.
\end{align}
The differential equations governing the dynamical evolution of the Universe can be obtained from equation (\ref{fRT field equations}) by choosing suitable line element. For cosmological applications one can use a spatially flat {\sf FLRW} metric given as
\begin{align}\label{metricFRW}
ds^{2}=-dt^{2}+a^{2}(t) \Big{(}dr^{2}+r^{2}d\Omega^2\Big{)},
\end{align}
where $a(t)$ is the scale factor of the universe and $d\Omega^2$ is the standard line element on a unit two sphere.
Applying metric (\ref{metricFRW}) to field equation (\ref{fRT field equations}) together with using the definition for $\mathbf{\Theta_{\mu \nu}}$ for a perfect fluid leads to
\begin{align}\label{first}
&&3H^{2}F({\sf R},{\sf T})+\frac{1}{2} \Big{(}f({\sf R},{\sf T})-F({\sf R},{\sf T}){\sf R}\Big{)}+3\dot{F}({\sf R},{\sf T})H\nn&&=
\Big{(}\kappa^{2} +{\mathcal F} ({\sf R},{\sf T})\Big{)}\rho+{\mathcal F} ({\sf R},{\sf T})p,
\end{align}
as the modified Friedmann equation, and
\begin{align}\label{second}
&2F({\sf R},{\sf T}) \dot{H}+\ddot{F} ({\sf R},{\sf T})-\dot{F} ({\sf R},{\sf T}) H=-\Big{(}\kappa^{2}+{\mathcal F} ({\sf R},{\sf T})\Big{)}(\rho+p),
\end{align}
as the modified Raychaudhuri equation, where $H$ indicates the
Hubble parameter. Note that, we have used ${\sf L}^{\sf{{\sf (m)}}}=p$ for a perfect fluid
in the expression (\ref{theta}). Applying the Bianchi identity to the field equation (\ref{fRT field equations}) gives following covariant equation 
\begin{align}\label{relation}
&(\kappa^{2} +\mathcal {F})\nabla^{\mu}{\sf T}_{\mu \nu}+\frac{1}{2}\mathcal {F}\nabla_{\mu}{\sf T}
+{\sf T}_{\mu \nu}\nabla^{\mu}\mathcal {F}-\nabla_{\nu}(p\mathcal{F})=0,
\end{align}
where we have dropped the argument of $\mathcal {F}({\sf R},{\sf T})$ for abbreviation. Equation (\ref{relation}) tells us that the conservation of {\sf EMT} is not generally respected in $f({\sf R},{\sf T})$ gravity theories. There is Only a narrow
class of solutions for which the conservation of energy is still preserved~\cite{Shabani172}. In this work we consider a specific class of models for which the Lagrangian is written as a minimal coupling between the Ricci curvature scalar and a function of trace of {\sf EMT}, i.e.,
\begin{align}\label{minimal}
f({\sf R},{\sf T})={\sf R}+\alpha\kappa^{2} h({\sf T}).
\end{align}
Also, we study bouncing solutions when the Universe contains a single perfect fluid with a barotropic {\sf EoS} given as $p=w\rho$. In the next section, we seek for non-singular cosmological solutions in $f({\sf R},{\sf T})$ gravity with a baratropic perfect fluid. As we shall see, a useful approach for choosing the functionality of $h({\sf T})$ is to rewriting equations (\ref{first}) and (\ref{second}) in terms of an effective fluid that respects the conservation of {\sf EMT}.
For the Lagrangian, (\ref{minimal}) equations (\ref{first}) and (\ref{second}) are simplified to
\begin{align}\label{eom1}
3H^{2}=\kappa^{2}\Bigg{\{}\Big{[}1+(1+w)\alpha h'\Big{]}\frac{{\sf T}}{3w-1}-\frac{\alpha h}{2}\Bigg{\}},
\end{align}
and
\begin{align}\label{eom2}
2\dot{H}=-\kappa^{2}\frac{w+1}{3w-1}(1+\alpha h'){\sf T},
\end{align}
where \lq\lq{}$\prime$\rq\rq{} denotes derivative with respect to the argument.
Also, from equation (\ref{relation}) we obtain
\begin{align}\label{relation-w}
\Big{(}1 + \frac{\alpha}{2}(3-w)h' + \alpha (1+w) Th''\Big{)}\dot{{\sf T}}+
3H(1+w)\Big{(}1 + \alpha h'\Big{)}{\sf T}=0.
\end{align}
As a matter of fact, one can directly solve equations (\ref{eom1})-(\ref{relation-w}) to obtain the scale factor or the Hubble parameter, once the $h({\sf T})$ function is determined. Alternatively, one can also define the pressure and energy density profiles of an effective fluid along with imposing the energy conditions in order to obtain the functionality of $h({\sf T})$. We then proceed in this way and rewrite equation (\ref{eom1}) as
\begin{align}\label{re1}
3H^{2}=\kappa^{2}\rho_{({\sf eff})}({\sf T}),
\end{align}
where,
\begin{align}\label{re2}
\rho_{({\sf eff})}({\sf T})\equiv \Big{[}1+(1+w)\alpha h'\Big{]}\frac{{\sf T}}{3w-1}-\frac{\alpha h}{2}.
\end{align}
With this definition the acceleration of expansion of the Universe can be obtained as follows
\begin{align}\label{re3}
\frac{\ddot{a}}{a}=-\frac{\kappa^{2}}{6}\Big{(}\rho_{({\sf eff})}({\sf T})+3p_{({\sf eff})}({\sf T})\Big{)},
\end{align}
where, we have defined the effective pressure as
\begin{align}\label{re4}
p_{({\sf eff})}({\sf T})\equiv\frac{w}{3w-1}{\sf T}+\frac{\alpha h}{2}.
\end{align}
Therefore, the original field equations of $f({\sf R},{\sf T})$ gravity can be recast into the usual Friedman form with an effective fluid. This fluid is characterized by an effective energy density and effective pressure which in turn are determined in terms of the {\sf EMT} trace. Therefore, once a property/relation for energy density and pressure components is established, a first order differential equation for the function $h({\sf T})$ could be reached. Solving the differential equation leads to an $h({\sf T})$ function that conveys the specified property/relation. Hence, in this way we can obtain a minimal $f({\sf R},{\sf T})$ model based on the conditions on energy density and pressure profiles of the effective fluid, instead of choosing the functionality of $f({\sf R},{\sf T})$ based on ad hoc mathematical terms. It is straightforward to verify that equation (\ref{relation-w}) is turned to the usual conservation equation in terms of $\rho_{(\sf{eff})}$ and $p_{(\sf{eff})}$, i.e.,
\begin{align}\label{re5}
\dot{\rho}_{({{\sf eff}})}+3H(\rho_{({{\sf eff}})}+p_{({{\sf eff}})})=0,
\end{align}
where the arguments are dropped for the sake of simplicity. Consequently, an effective {\sf EoS} parameter can be defined for this effective fluid as
\begin{align}\label{re6}
\mathcal{W}\equiv \frac{p_{({\sf eff})}}{\rho_{({\sf eff})}}=-1+\frac{2(1+w)(1+\alpha h'){\sf T}}{2\left[1+(1+w)\alpha h'\right]{\sf T}-(3w-1)\alpha h}.
\end{align}
\section{Asymptotic de-Sitter bouncing solutions in $f({\sf R},{\sf T})$ gravity}\label{Sec2}
In this section we study different bouncing cosmological solutions of $f({\sf R},{\sf T})$ gravity. We extract those solutions which correspond to some properties of the effective fluid. Such an approach may help us to understand how these solutions can emerge in $f({\sf R},{\sf T})$ gravity. Furthermore, there can be obtained more bouncing solutions, however, to show that $f({\sf R},{\sf T})$ gravity theories are capable of describing a non-singular pre-Big Bang era, we are restrict ourselves to study only few examples. We set $\kappa^{2}=1$ in the rest of the work. 

\subsection{Type {\sf A} models: Solutions which correspond to a constant effective pressure, $p_{({\sf eff})}({\sf T})=\mathcal{P}$}\label{Sec21}
\begin{figure}[h!]
\centerline{\includegraphics[width=.65\textwidth,trim=10 0 0 0]{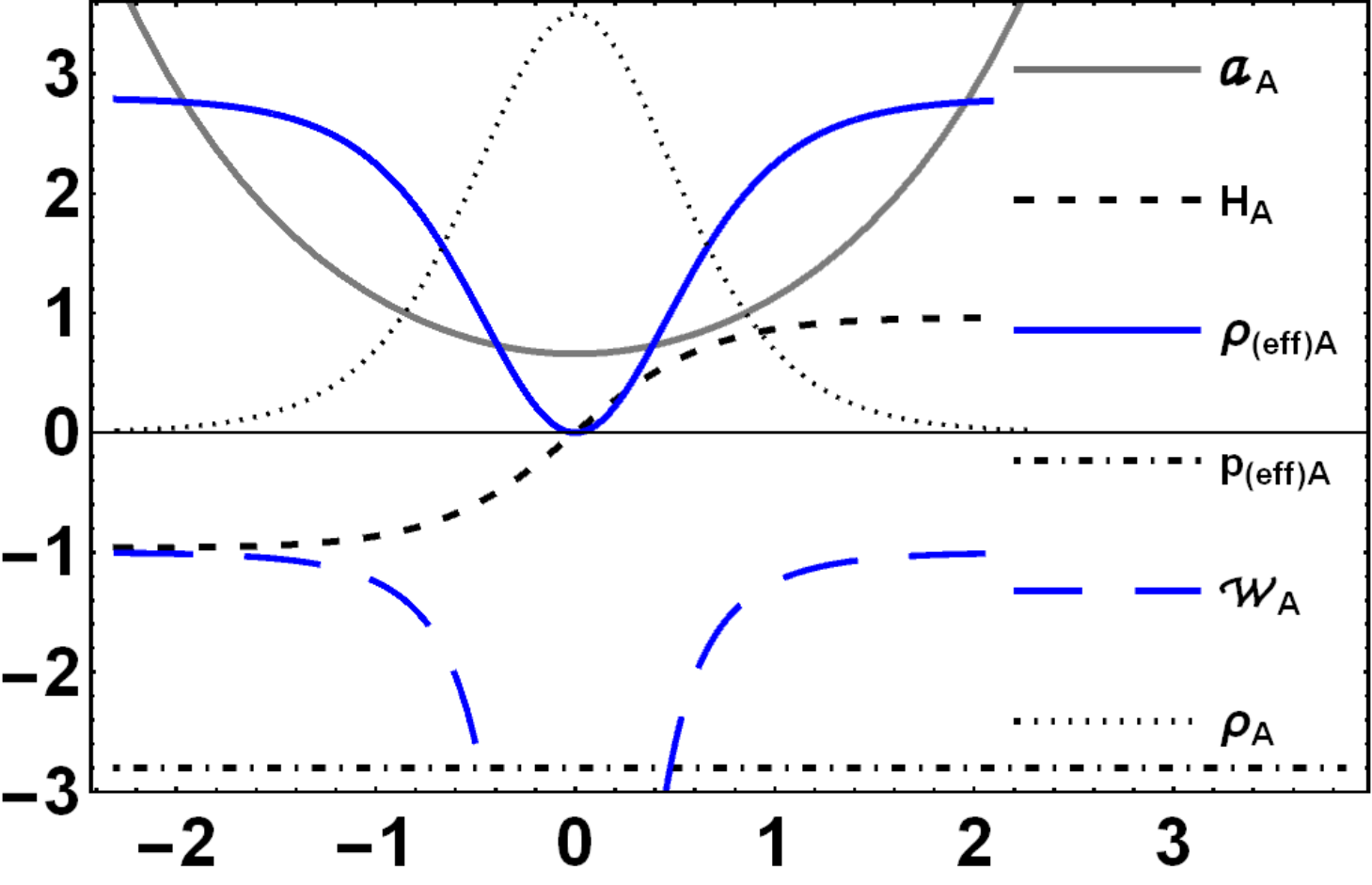}}
\caption{\label{figA1} Different cosmological quantities related to the model {\sf A} which are characterized by solution (\ref{e214}). The black solid line indicates the scale factor. The black middle sized dashed line shows the Hubble parameter, the solid blue curve indicates the effective energy density, the black dot-dashed curve shows the pressure, The long-dashed blue curve shows the effective {\sf EoS} and the black dotted curve shows the effective energy density. We see that $\mathcal{W}_{\sf{A}}<-1$ always, signaling that the scalar representation of model {\sf A} should be defined by a phantom field. We have set $w=0.6$, $\rho_0=1$, ${\mathcal P}=-2.8$ and $a_0=1.12.$ The horizontal axis represents the cosmic time.}
\end{figure}

Let us begin with bouncing solutions which are obtained by assuming an effective fluid with constant pressure. We show that these type of models lead to a de-Sitter era at late times~\cite{Shabani18}. From definition (\ref{re4}) for a constant effective pressure we obtain
\begin{align}\label{e211}
h_{{\sf A}}({\sf T})=\frac{2}{\alpha }\left(\mathcal{P}+\frac{w}{1-3 w}{\sf T}\right).
\end{align}
Substituting the above function into the modified conservation equation (\ref{relation-w}) together with using ${\sf T}=(3w-1)\rho_{{\sf A}}$ for a perfect fluid we get
\begin{align}\label{e212}
\rho_{{\sf A}}=\rho_{0}a_{{\sf A}}^{-3}.
\end{align}
where $\rho_{0}$ is an integration constant. Substituting solutions (\ref{e211}) and (\ref{e212}) back into the Friedman equation (\ref{re1}) leaves us with the following differential equation for the scale factor
\begin{align}\label{e213}
3\left(\frac{a'_{{\sf A}}(t)}{a_{{\sf A}}(t)}\right)^{2}-\frac{\rho_{0} \left(w^2-1\right)}{(3 w-1) a_{{\sf A}}(t)^3}+\mathcal{P}=0.
\end{align}
For $\mathcal{P}=0$, a particular solution to the above equation is found as $a_{{\sf A}}(t)\propto t^{2/3}$. Moreover, in case in which the middle term is absent, the solution represents a de-Sitter phase for $\mathcal{P}<0$.  From the Friedman equation (\ref{e213}), we observe that the effective pressure $\mathcal{P}$ plays the role of a cosmological constant. Other forms for the Friedman equation similar to equation (\ref{e213}) have been found in the literature. For example in~\cite{Babichev05} authors have worked on a perfect fluid, dubbed as {\sf DE} with a linear {\sf EoS}, in a flat {\sf FLRW} background. They used $p_{{\sf DE}}=\alpha(\rho_{{\sf DE}}-\rho_{0})\equiv p_{\alpha}+p_{\Lambda}$, where $\alpha$ and $\rho_{0}$ being some constants. The {\sf DE} pressure is then decomposed to a dynamical part which corresponds to $p_{\alpha}$ as well as a constant part which corresponds to $p_{\Lambda}$. These assumptions lead to an equation that will reduce to (\ref{e213}), when only $p_\Lambda$ is considered. In the upcoming subsections when we study the model {\sf C}, we find that in the context of $f({\sf R},{\sf T})$ gravity one can obtain the same results as those given in~\cite{Babichev05}. Equation (\ref{e213}) admits three different general solutions which we will consider only one of them. A general solution of equation (\ref{e213}) is obtained as
\begin{align}\label{e214}
&a_{{\sf A}}(t)=\frac{1}{\mathfrak{A}}\left[\cosh \left(\sqrt{-\frac{\mathcal{P}}{3}t}\right)-\sinh \left(\sqrt{-\frac{\mathcal{P}}{3}t}\right)\right]\nonumber\\
&&\hspace{-9cm}\times\left\{\mathfrak{B}+\frac{(w^2-1)\mathcal{P}\rho_{0}}{2(3w-1)}\left[\sinh \left(\sqrt{-3\mathcal{P}} t\right)+\cosh \left(\sqrt{-3\mathcal{P}} t\right)-1\right]\right\}^{2/3}\nonumber\\
&\mathfrak{A}=\left[\mathcal{P}^2 \left(2\mathfrak{B}-\frac{\left(w^2-1\right)\mathcal{P} \rho _0}{3 w-1}\right)\right]^{1/3}\nonumber\\
&\mathfrak{B}=a_{0}^3 \mathcal{P}^2-\sqrt{a_{0}^3 \mathcal{P}^3 \left(a_{0}^3 \mathcal{P}+\frac{\rho_{0}(1- w^2)}{3 w-1}\right)}.
\end{align}
Note that once we set the integration constant in equation (\ref{e213}) such that $a_{\textrm{A}}(0)=0$, we get $\mathfrak{B}=0$ and furthermore, if we consider the case $w=0$, we obtain $\mathfrak{A}=\mathcal{P} \rho_{0}^{1/3}$. Thus the solution (\ref{e214}) reduces to the familiar form $a(t)=(-\rho_{0}/\mathcal{P})^{1/3} \sinh ^{2/3}(\sqrt{-3\mathcal{P}/2}t)$. However, the integration constant in solution (\ref{e214}) is fixed such that we have $a(t=0)=a_{0}$. The most important feature of solution (\ref{e214}) is the appearance of cosine hyperbolic function which allows us to have a non-singular behavior for $\mathcal{P}<0$. This solution describes the de-Sitter expansion of the Universe which is initially dominated by dust~\cite{Frieman08}. In Figure~\ref{figA1} the thick black curve shows the scale factor solution (\ref{e214}). This solution behaves exponentially in the far past and far future from the bounce and reaches a nonzero minimum value at the bounce time, given by
\begin{align}\label{e219}
a_{\sf{(min)A}}=\mathfrak{A}^{-1}\sqrt[3]{\frac{2(1-3w)\mathfrak{B}}{(1- w^2)\mathcal{P} \rho_{0}}-1} \left[\frac{2  (3 w-1)\mathfrak{B}^2}{2 (3 w-1)\mathfrak{B}-\left(w^2-1\right)\mathcal{P} \rho_{0}}-\right.
\left.\frac{\left(w^2-1\right)\mathcal{P} \rho_{0}}{3 w-1}\right]^{2/3}.
\end{align}
The Hubble parameter is obtained as
\begin{align}\label{e215}
H_{\sf{A}}(t)=\sqrt{-\frac{\mathcal{P}}{3}}\left[\frac{2 (1-3 w)\mathfrak{B}+\left(w^2-1\right)\mathcal{P} \rho_{0}\left(\sinh \left(\sqrt{-3\mathcal{P}} t\right)+\cosh \left(\sqrt{-3\mathcal{P}} t\right)+1\right)}{2 (3 w-1)\mathfrak{B}+\left(w^2-1\right)\mathcal{P} \rho_{0} \left(\sinh \left(\sqrt{-3\mathcal{P}} t\right)+\cosh \left(\sqrt{-3\mathcal{P}} t\right)-1\right)}\right].
\end{align}
Figure~\ref{figA1} shows that the Hubble parameter tends to constant values before and after the bounce and vanishes at the bounce. On the other hand, the Hubble parameter and its time derivatives diverge at an imaginary time $t_{\sf{s}}=i\pi/\sqrt{-3\mathcal{P}}-t_{\sf{b}}$, where $t_{\sf{b}}$ is the (real) time at which the bounce occurs. Therefore, we observe that the Hubble parameter behaves in a well-defined way (without any future singularity) so that it respects the bounce conditions. We can also obtain the effective energy density and the effective {\sf EoS} parameter. To this aim, we substitute (\ref{e211}) into equations (\ref{re2}) and (\ref{re6}), respectively. Thus we obtain
\begin{align}
&\rho_{({\sf eff)A}}(a)=\frac{\left(w^2-1\right)\rho_{0}}{(3 w-1)a_{\sf{A}}^3}-\mathcal{P}\label{e216},\\
&\mathcal{W}_{\sf{A}}=\frac{(1-3 w)\mathcal{P}a_{\sf{A}}^3}{(3 w-1)\mathcal{P}a_{\sf{A}}^3-\left(w^2-1\right)\rho_{0}}\label{e217}.
\end{align}
~
As can be seen from Figure~\ref{figA1}, the effective density reduces from a constant value and tends to zero near the bounce. From equation (\ref{re1}) we see that the vanishing of Hubble parameter at the bounce demands that the effective density becomes zero. Also for the same reason, the effective {\sf EoS} diverges at the bounce. Such behaviors are common for all bouncing models that we shall present in the framework of minimal $f({\sf R},{\sf T})$ gravity. the matter energy density itself increases from small values to a maximum value near the bounce. Based on the exchange of energy between gravitational field and matter constituents (that the mechanism of which is explained in~\cite{Harko14}) one may explain the process of bouncing behavior; the interaction of the real fluid with curvature leads to some transformations of energy from gravitational field to matter before the bounce where the spacetime curvature is dominant in comparison to matter energy density. Such a transmutation, that the start of which is triggered at the time far past the bounce, gives rise to an increase in the energy density as the bounce event is approached. At the bounce time the energy density of matter grows to a maximum value after which the process of transmutation is reversed until the density falls back to zero (the post-bounce regime). Note that the effective energy density remains constant in the de-Sitter era. However, some physics is needed in order to explain the process of matter production from curvature component which disturbs the stability of de-Sitter era to enter the bounce event. 

Informations from Figure~\ref{figA1} can help us to discuss the energy conditions. In {\sf GR} the well-known energy conditions are the {\sf NEC}, {\sf WEC}, {\sf SEC} and the dominant energy condition ({\sf DEC}). In a modified gravity theory with defined effective energy density and pressure, these conditions can be written as~\cite{Sharif13}
\begin{align}
&{\sf WEC} \Leftrightarrow \rho_{({\sf eff})}\geq0,~\mbox{and}~\rho_{{\sf(eff)}}+p_{{\sf(eff)}}\geq0,\label{wec}\\
&{\sf NEC} \Leftrightarrow  \rho_{{\sf(eff)}}+p_{{\sf(eff)}}\geq0,\label{nec}\\
&{\sf SEC} \Leftrightarrow   \rho_{{\sf(eff)}}+3p_{{\sf(eff)}}\geq0~ \mbox{and}~\rho_{{\sf(eff)}}+p_{{\sf(eff)}}\geq0,\label{sec}\\
&{\sf DEC} \Leftrightarrow \rho_{{\sf (eff)}}\geq0,~\mbox{and}~\rho_{\sf{(eff)}}\pm p_{\sf{(eff)}}\geq0.\label{dec}
\end{align}
For the model {\sf A}, we obtain the following results
\begin{align}\label{e2171}
&\rho_{\sf{(eff)}}+p_{\sf{(eff)}}=\frac{\left(w^2-1\right)\rho_{0}}{(3 w-1)a_{\sf{A}}^3},\\
&\rho_{\sf{(eff)}}+3p_{\sf{(eff)}}=\frac{\left(w^2-1\right)\rho_{0}}{(3 w-1)a_{\sf{A}}^3}+2\mathcal{P}.
\end{align}
It is obvious that the fulfillment of {\sf NEC} and thus {\sf WEC} requires $(w^{2}-1)/(3w-1)\geq0$ which gives $-1\leq w \leq\frac{1}{3}$. From (\ref{e216}) we see that the effective energy density tends to $-\mathcal{P}$ at late times and vanishes at the time of bounce. Therefore, we obtain $3 \mathcal{P}\leq\rho_{\textrm{(eff)}}+3p_{\textrm{(eff)}}\leq2\mathcal{P}$, (remember that $\mathcal{P}<0$). Thus, since only negative values are valid for the effective pressure $\mathcal{P}$, the {\sf SEC} is always violated. However, the validity of {\sf NEC} depends upon the value of $w$. We plot the diagrams for $w=0.6$ in Figure~\ref{figA1}. This figure shows that the {\sf NEC} and {\sf SEC} are violated in this case. Our studies show that the bouncing behavior is achieved from solution (\ref{e214}) for $w>1/3$. Note that, as we have mentioned before, solution (\ref{e214}) is only one of the three possible solutions of equation (\ref{e213}). Investigating other solutions may validate the cases with $w<1/3$ from energy conditions point of view.

\begin{figure}[h!]
\centering 
\centerline{\includegraphics[width=.65\textwidth,trim=10 0 -3 0]{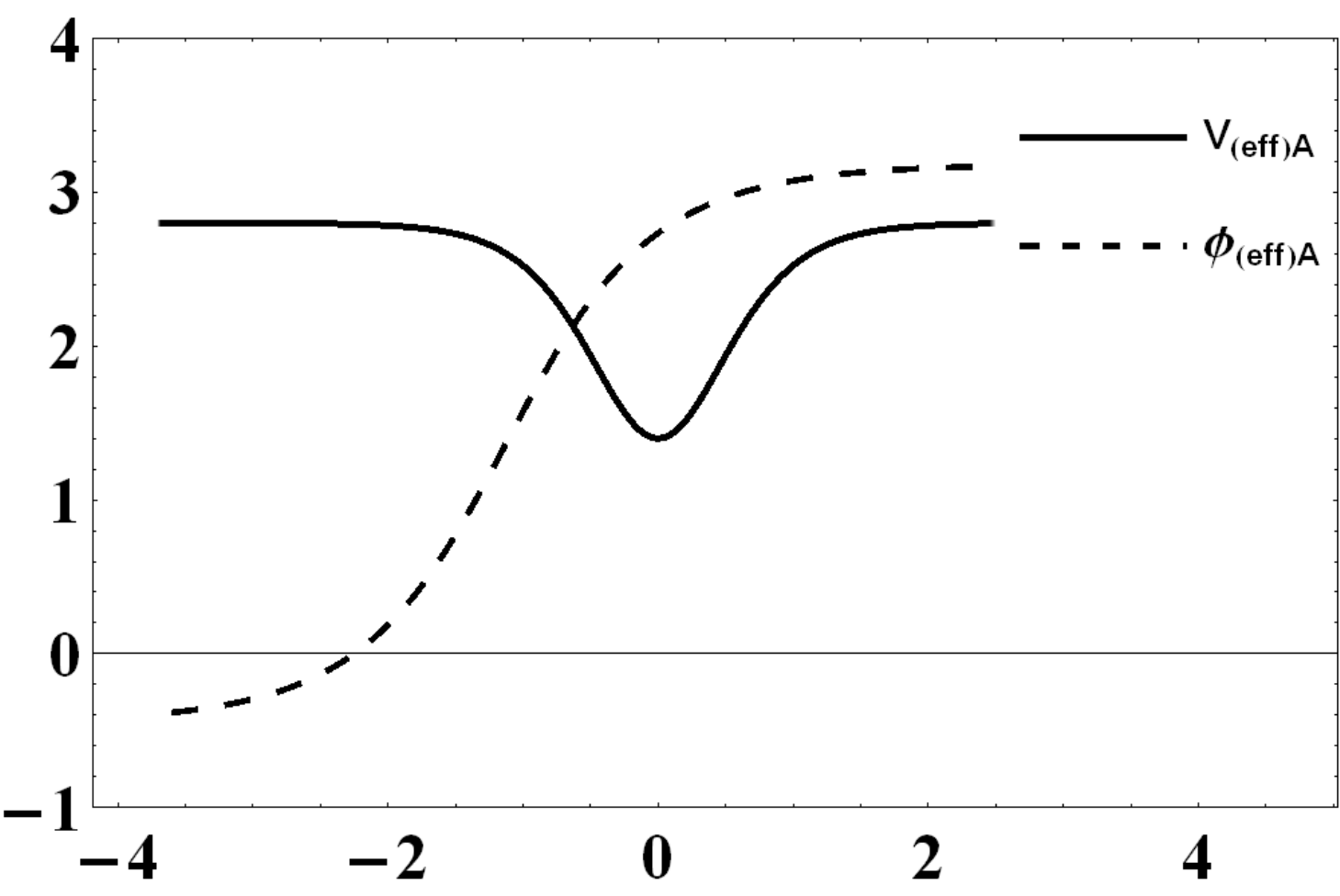}}
\caption{\label{figA2} The phantom scalar field representation of model {\sf A}. Solid curve shows the effective potential and the dashed one indicates the phantom scalar field.}
\end{figure}

\begin{figure}[h!]
\centering 
\centerline{\includegraphics[width=.65\textwidth,trim=10 0 -2 0]{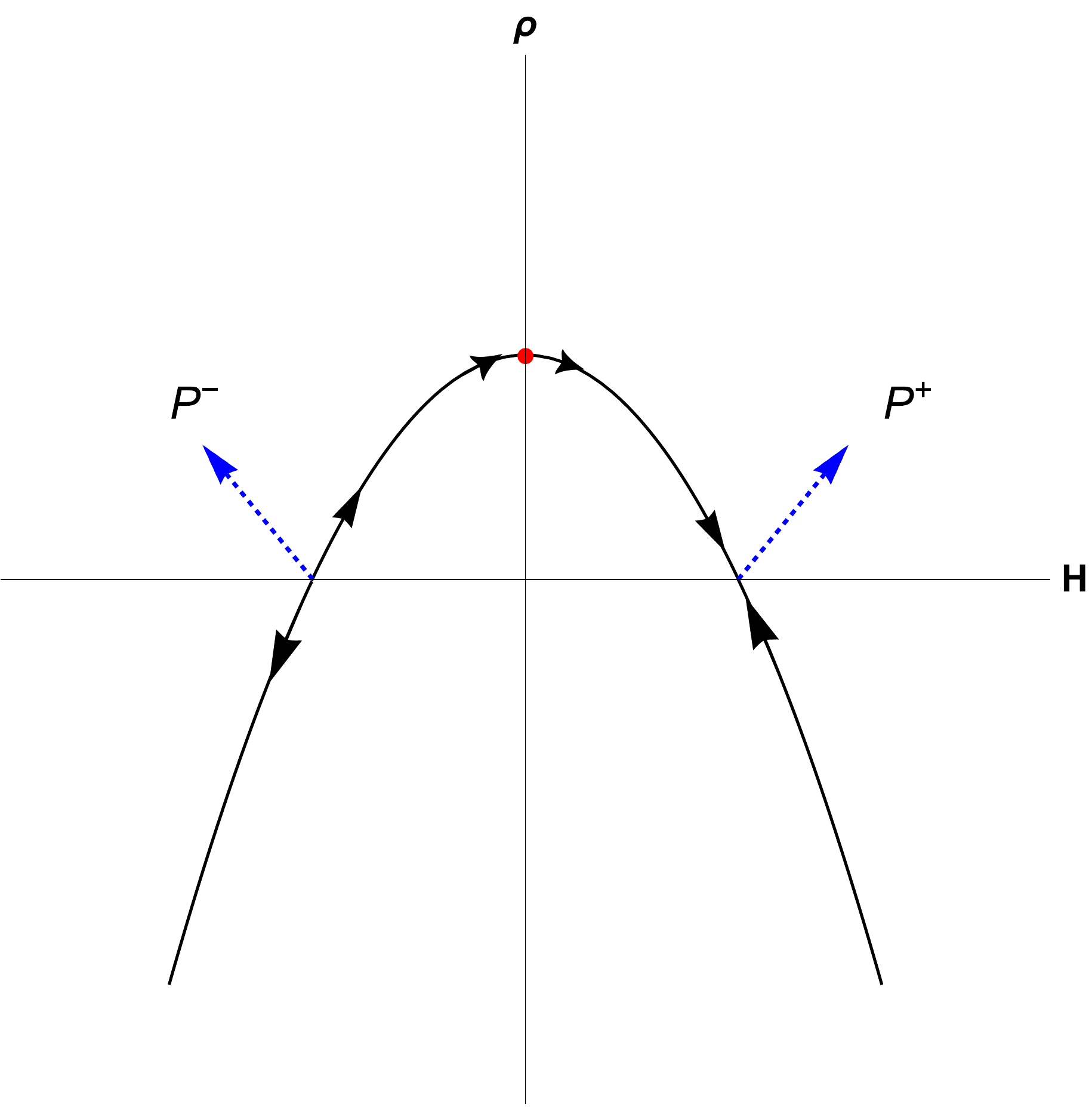}}
\caption{\label{figA3}A typical phase trajectory in the phase space plane $(H,\rho)$ corresponding to model {\sf A}. This plot shows that the evolution of the Universe has started from an unstable state and terminates in a stable phase. The bounce event is an unstable (of saddle type) fixed red point.}
\end{figure}

One may be tempted to reinterpret the source of matter as that of a scalar field. Such a representation is also used in similar works~\cite{Barrow90,Contreras16,Contreras17, Chavanis13}. In the case of constant effective pressure, the particular solution (\ref{e214}) (which is valid for $1/3<w<1$) corresponds to ${\mathcal W}_{\sf{A}}<-1$ which can be realized from relation (\ref{e217}), see also the long-dashed blue curve in Figure~\ref{figA1}. Therefore, if we want to translate mutual interactions of perfect fluid and curvature as the behavior of an effective scalar field, we should employ a phantom scalar field. Therefore, we can define
\begin{align}\label{e222}
&\mathcal{L}_{\sf{Ph}}=-\frac{1}{2}{\phi_{\sf{(eff)}}}_{;\mu}{\phi_{\sf{(eff)}}}^{;\mu}-V(\phi_{\sf{(eff)}}),\\
&\rho_{\sf{(eff)}}=-\frac{1}{2}\dot{\phi}_{\sf{(eff)}}^2+V(\phi_{\sf{(eff)}}),~~~~~~~~~p_{\sf{(eff)}}=-\frac{1}{2}\dot{\phi}_{\sf{(eff)}}^2-V(\phi_{\sf{(eff)}}).
\end{align}
where, the subscript ${\sf Ph}$ denotes the phantom field and \lq\lq{};\rq\rq{} indicates covariant differentiation. We then get
\begin{align}\label{e223}
&\dot{\phi}_{\sf{(eff)A}}^2=-(\rho_{\sf{(eff)}}+p_{\sf{(eff)}})=\frac{\left(1-w^2\right)\rho_{0}}{(3 w-1)a_{\sf{A}}^3},\\
&V(\phi_{\sf{(eff)A}})=\frac{1}{2}\left(\rho_{\sf{(eff)}}-p_{\sf{(eff)}}\right)=\frac{\left(w^2-1\right)\rho_{0}}{2(3 w-1)a_{\sf{A}}^3}-\mathcal{P}.
\end{align}
A straightforward calculation reveals that
\begin{align}\label{e219}
\phi_{\sf{(eff)A}}=-4\frac{\mathfrak{A}}{\mathcal{P}} \sqrt{\frac{(1 - 3 w) (w-1)\mathfrak{A}}{3\mathfrak{C}}}
\arctan\left[\frac{\mathcal{B} (3 w-1) \left(\tanh \left(\sqrt{\frac{-3p}{16}}t\right)-1\right)+ \left(w^2-1\right)\mathcal{P},\rho_{0}}{\sqrt{- (w+1)\mathcal{P}\rho_{0} \mathfrak{C}}}\right],
\end{align}
where we have defined
\begin{align}\label{e220}
 \mathfrak{C}=(w-1) \Big[2(1-3 w)\mathfrak{B}+\left(w^2-1\right)\mathcal{P}\rho_{0}\Big].
\end{align}
We can also obtain $\phi_{\sf{(eff)A}}$ in terms of the scale factor by solving solution (\ref{e214}) for time $t$, which gives
\begin{align}\label{e221}
t=\frac{1}{\sqrt{-3\mathcal{P}}}\log\left[\frac{2 a_{\sf{A}}^3 \mathfrak{A}^3 (1-3 w)^2 \left(1-\sqrt{\frac{\mathfrak{C} (w+1)\mathcal{P}\rho_{0}}{a_{\sf{A}}^3 \mathfrak{A}^3 (1-3 w)^2}+1}\right)+ (w+1)\mathcal{P}\rho_{0}\mathfrak{C}}{\left(w^2-1\right)^2\mathcal{P}^2\rho_{0}^2 }\right].
\end{align}
We have plotted $\phi_{\sf{(eff)A}}$ and $V_{\sf{(eff)A}}$ in Figure~\ref{figA2} for the same parameters of Figure~\ref{figA1}.
Another important issue that needs to be treated is to investigate the stability properties of solution of equation (\ref{e213}). Substituting solution (\ref{e216}) in (\ref{re1}) and also (\ref{e212}) in (\ref{eom2}) and taking $H$ and $\rho$ as dynamical variables, we arrive at the following dynamical system
\begin{align}
&\dot{H}=\frac{\beta}{2} \rho,\label{e224}\\
&\dot{\rho}=-3 H \rho,\label{e225}\\
&3H^{2}=-\beta \rho-\mathcal{P}\label{e226},
\end{align}
where $\beta=(1-w^{2})/2(3w-1)$ and we have dropped subscripted ${\sf A}$ . Note that the validity of solution (\ref{e214}) requires that $\beta>0$. The System (\ref{e224})-(\ref{e226}) has two critical points far from the bounce; ${\sf P}^{(\pm)}=(\pm\sqrt{-\mathcal{P}/3},0)$ which corresponds to the eigenvalues $\lambda^{(\pm)}=(\mp\sqrt{-\mathcal{P}/3},0)$. We see that the stability properties of solutions are independent of $w$. Due to the appearance of the zero eigenvalues, one may not decide about the stability properties of these fixed points, however, by inspecting equations (\ref{e224}) and (\ref{e225}) it is possible to figure out the nature of the fixed points. For fixed point ${\sf P}^{(-)}$ equation (\ref{e225}) becomes
\begin{align}\label{e227}
\dot{\rho}^{(-)}=\sqrt{-3\mathcal{P}} \rho^{(-)}.
\end{align}
Therefore, in the vicinity of ${\sf P}^{(-)}$ within the phase space, by indicating the values of $\rho$ and $H$ on vertical and horizontal axises, respectively, we have $\dot{H}>0~\&~\dot{\rho}>0$ for all points with $\rho>0$ and $\dot{H}<0~\&~ \dot{\rho}<0$ for points with $\rho<0$. These show that when $\rho\rightarrow\rho+\delta\rho$ for $t\rightarrow t+\delta t$, the solution at ${\sf P}^{(-)}$ will not stay stationary and hence it is a repulsive fixed point. On the other hand, for ${\sf P}^{(+)}$ we have
\begin{align}\label{e228}
\dot{\rho}^{(+)}=-\sqrt{-3\mathcal{P}} \rho^{(+)}.
\end{align}
Therefore, in this case we have $\dot{H}>0~\&~\dot{\rho}<0$ for all points with $\rho>0$ and $\dot{H}<0~\&~ \dot{\rho}>0$ for points with $\rho<0$ in the vicinity of ${\sf P}^{(+)}$. Hence it is a stable fixed point. The bounce corresponds to the point ${\sf P}^{\textrm{(b)}}=(0,-\mathcal{P}/\beta)$ for which we have $\dot{H}=-\mathcal{P}/2>0$. Therefore, at this point the tangent vectors on the phase space trajectories are directed toward the right side. We plot a typical trajectory in the phase space in Figure~\ref{figA3}.
\subsection{Solutions which correspond to a general effective {\sf EoS}, $p_{(\sf{eff})}=\mathcal{Y}(\rho_{(\sf{eff})})$}\label{Sec22}
These class of models can be constructed by imposing a particular condition on the effective profiles. This approach can be viewed as a sort of classification of $f({\sf R},{\sf T})$ gravity models based on the properties of the effective quantities. Generally, one can obtain a class of $h({\sf T})$ functions for a determined property which is specified by an effective {\sf EoS}. In the following sections we consider two subclasses based on conditions on the effective densities. We find that each class of $h({\sf T})$ solutions that exhibit bouncing behavior correspond to an effective {\sf EoS} which is already introduced or obtained for an exotic fluid in the literature~\cite{Contreras17,Contreras16,Babichev05,Chavanis13,Bamba12}.

\subsubsection{Type {\sf B} Models: Solutions which follow the relation $d\rho_{\sf{(eff)}}/d{\sf T}=[n/(1+w){\sf T}](\rho_{\sf{(eff)}}+p_{\sf{(eff)}})$ }\label{sec231}

\begin{figure}[h!]
\centering 
\centerline{\includegraphics[width=.65\textwidth,trim=10 0 0 0]{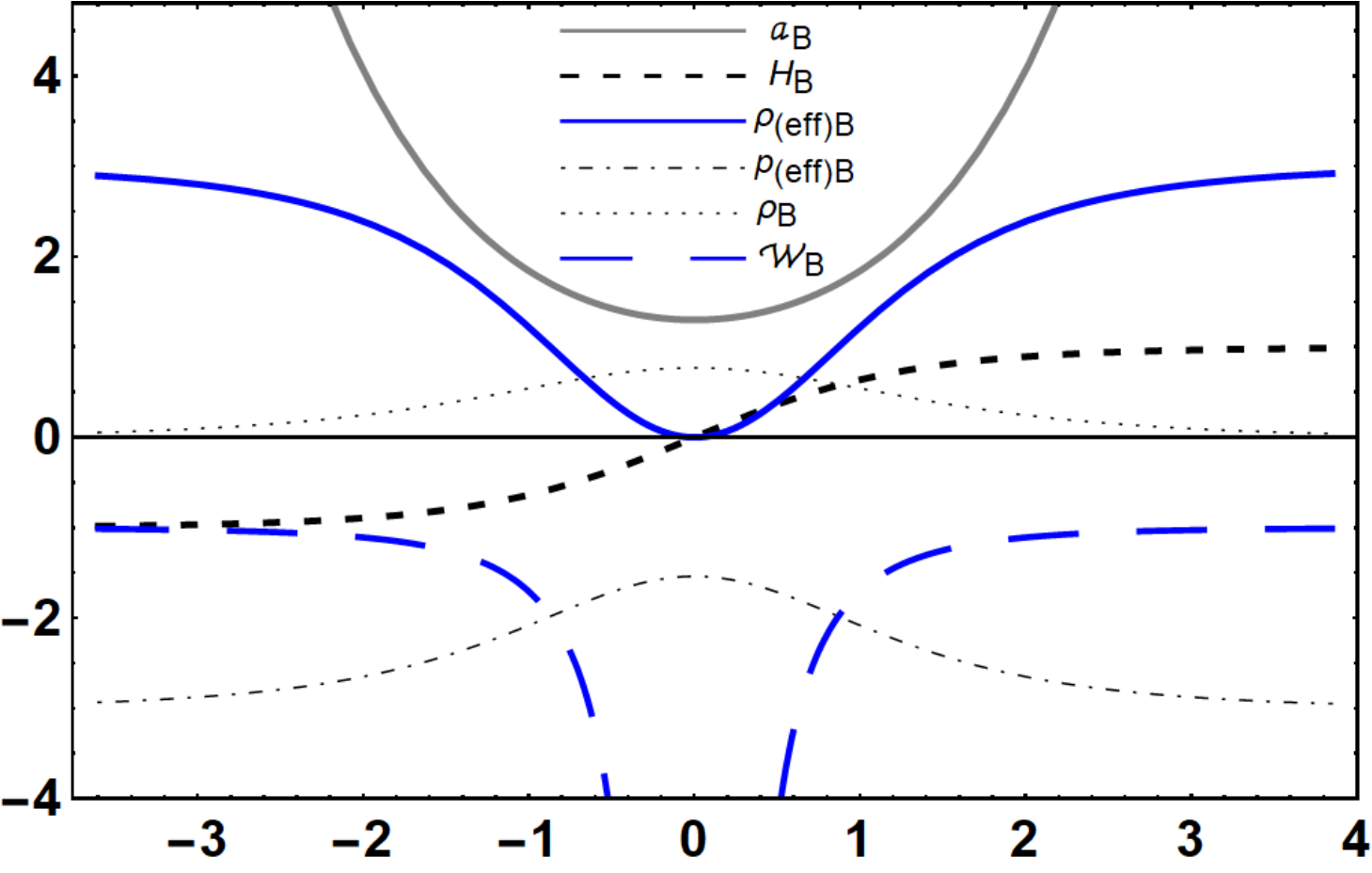}}
\caption{\label{figC}Cosmological parameters of the bouncing solution for the model {\sf B}. We have set $\alpha= 20$, $\Lambda_{{\sf B}}= -0.3$ and $\rho_0 = 1$.}
\end{figure}

Applying this condition together with using the definitions for the effective energy density and pressure, we arrive at a differential equation for $h({\sf T})$ which can be solved as follows~\cite{Shabani172}
\begin{align}\label{e231}
h_{\sf{B}}({\sf T})=\frac{2 \Gamma_{\sf{B}} (w+1)}{2 n+3 w-1} {\sf T}^{\frac{2 n+3 w-1}{2 (w+1)}}-\frac{2 (n-1) }{\alpha  (2 n+w-3)}{\sf T}+\Lambda_{\sf{B}},
\end{align}
where, $\Gamma_{\sf{B}}$ and $\Lambda_{\sf{B}}$ are integration constants and $n$ is an arbitrary constant. Substituting the relation $d\rho_{\sf{eff}}/d{\sf T}=[n/(1+w){\sf T}](\rho_{\sf{eff}}+p_{\sf{eff}})$ into equation (\ref{relation-w}) gives
\begin{align}\label{e232}
\rho_{\textrm{\sf{B}}}=\rho_{0}a^{-\frac{3 (w+1)}{n}}.
\end{align}
Next we proceed to find a non-singular bouncing solution by solving the modified Friedmann equation (\ref{eom1}) for solutions (\ref{e231}) and (\ref{e232}). We first try to obtain solutions of the form $a_{{\sf B}}(t)=\mathcal{R}(\cosh[(t-t_{0})/\mathcal{R}]-\mathcal{S})$ where $\mathcal{R}$ and $\mathcal{S}$ are constants. This type of solution has been discussed in~\cite{Contreras17} under assumption of an {\sf MEoS} and has the following form of the  Friedmann equation
\begin{align}\label{e233}
3\left(\frac{a'_{\sf{B}}(t)}{a_{\sf{B}}(t)}\right)^{2}-\frac{3(\mathcal{S}^{2}-1)}{a_{\sf{B}}(t)^{2}}-\frac{6\mathcal{S}}{\mathcal{R}a_{\sf{B}}(t)}-\frac{3}{\mathcal{R}^{2}}=0.
\end{align}
Applying (\ref{e231}) and (\ref{e232}) in equation (\ref{eom1}) gives the Friedmann equation for arbitrary constants $w$ and $n$. We can check that, there are only two cases which correspond to equation (\ref{e233}) and thus to the scale factor $a_{\sf{B}}$ as the solution; when $w=-1/5,n=12/5$ and $w=-1/5,n=6/5$. However, the latter leads to similar physics to the former. The physical quantities constructed out of the bouncing solution for $w=-1/5,~n=12/5$ are given as follows
\begin{align}
&a_{\sf{B}}(t)=\mathcal{R}\left[\cosh\left(\frac{t}{\mathcal{R}}\right)-\mathcal{S}\right],~~
\mathcal{R}=\sqrt{-\frac{6}{\alpha\Lambda_{\sf{B}}}},~~\mathcal{S}=-\frac{3}{10}\mathcal{R}\rho{0},\label{e234}\\
&H_{\sf{B}}\left(t\right)=\frac{\sinh \left(\frac{t}{\mathcal{R}}\right)}{\mathcal{R}\Big[\cosh \left(\frac{t}{\mathcal{R}}\right)-\mathcal{S}\Big]},\label{e235}\\
&\rho_{\sf{(eff)B}}=\frac{3 \Big(a_{\sf{B}}+\mathcal{R}(\mathcal{S}-1)\Big) \Big(a_{\sf{B}}+\mathcal{R}(\mathcal{S}+1)\Big)}{a_{\sf{B}}^2 \mathcal{R}^2},\label{e236}\\
&p_{\sf{(eff)B}}=\frac{\mathcal{R}\Big [\mathcal{R}-\mathcal{S} (4 a_{\sf{B}}+\mathcal{S} \mathcal{R})\Big]-3 a_{\sf{B}}^2}{a_{\sf{B}}^2 \mathcal{R}^2},\label{e237}\\
&\mathcal{W}_{\sf{B}}=-\frac{1}{3} \left[\frac{a_{\sf{B}}}{a_{\sf{B}}+\mathcal{R}(\mathcal{S}-1)}+\frac{a_{\sf{B}}}{a_{\sf{B}}+\mathcal{R}(\mathcal{S}+1)}+1\right].\label{e238}
\end{align}
The behavior of the above quantities are depicted in Figure~\ref{figC}. In order that the {\it ansatz} $a_{{\sf B}}(t)$ satisfies equation (\ref{e233}) we must have
\begin{align}\label{e239}
\Gamma_{\sf{B}}=\frac{1}{32 \alpha^{2}}\left( \frac{27}{\Lambda_{\sf{B}}}+\frac{50 \alpha }{\rho_{0}^2}\right).
\end{align}
Solution (\ref{e234}) shows that the Universe shrinks from an infinite size to a minimum radius which is equal to $\mathcal{R}(1-\mathcal{S})$. The size of Universe at the time of bounce is controlled by the constant $\Lambda_{\sf{B}}$ as well as the coupling constant $\alpha$.  As can be seen, in equation (\ref{e233}) and the subsequent solutions, these two constants appear as a multiplied form. This means that if either of these constants become zero, the bouncing solution would disappear. Also, expression (\ref{minimal}) together with solution (\ref{e231}) indicate that the bouncing solution disappears in model {\sf B} when $\alpha=0$. 
\par
The expressions for effective energy density and pressure, i.e. (\ref{e236}) and (\ref{e237}), can be rewritten as a sum of three densities and pressures given by
\begin{align}\label{e239-1}
&\rho_{\textrm{(eff)B}}=\frac{1-\mathcal{S}^2}{a_{\sf{B}}^2}-\frac{4\mathcal{S}}{a_{\sf{B}}\mathcal{R}}-\frac{3}{\mathcal{S}^2}=\rho_{1}+\rho_{2}+\rho_{3},\\
&p_{\sf{(eff)B}}=\frac{\mathcal{S}^2-1}{a_{\sf{B}}^2}+\frac{6\mathcal{S}}{a_{\sf{B}}\mathcal{R}}+\frac{3}{\mathcal{S}^2}=p_{1}+p_{2}+p_{3},
\end{align}
From these expressions, we could suppose that our effective fluid consists of a combination of three perfect fluids
with {\sf EoS}s $w_{i}=\rho_{i}/p_{i}$. In this view, there is a {\sf DE} component which corresponds to a cosmological constant with $w_{1}=-1$, a quintessence with $w_{2}=-2/3$ and a fluid which drives an expanding Universe with zero acceleration, with $w_{2}=-1/3$. Such a description has been presented in~\cite{Contreras17}. In the framework of $f(R,{\sf T})$ gravity this decomposition may be translated as follows; the effects of unusual interactions of matter (here a perfect fluid with $w=-1/5$) with curvature, could produce the same behavior as the case of {\sf GR} with three different fluids.
Eliminating the scale factor from solutions (\ref{e236}) and (\ref{e237}) leads to an effective {\sf EoS} of the form
\begin{align}\label{e239-1}
p_{\sf{(eff)B}}=-\frac{\rho_{\sf{(eff)B}}}{3}+\frac{2 \mathcal{S}}{3\mathcal{R}^2 \left(\mathcal{S}^2-1\right) }\left[3\mathcal{R}^2 \left(\mathcal{S}^2-1\right)\rho_{\sf{(eff)B}}+9\right]^{\frac{1}{2}}+\f{2}{\mathcal{R}^2\left(\mathcal{S}^2-1\right)}.
\end{align}
In the cosmological applications, a general type of {\sf EoS} $p=\beta\rho+\gamma f(\rho)$  is ascribed to some exotic or dark fluid which can determine the evolution of the Universe. Different choices for function $f(\rho)$ are studied in the literature. Such an equation has been already discussed in~\cite{Bamba12} where the authors considered cosmological consequences of an {\sf EoS} of the form $p=-\rho-\rho^{q}+1$.

Substituting expressions (\ref{e236}) and (\ref{e237}) in the {\sf NEC} and {\sf SEC} conditions (\ref{nec}) and (\ref{sec}) leads to the following conditions
\begin{align}
&{\sf NEC}_{\sf{(eff)B}}=\frac{2 \left(\mathcal{S}^2-1\right)}{a_{\sf{B}}^2}+\frac{2 \mathcal{S}}{a_{\sf{B}} \mathcal{R}}\geq0,\label{e239-2}\\
&{\sf SEC}_{\sf{(eff)B}}=-6\left(\frac{1}{\mathcal{R}^{2}}+\frac{\mathcal{S}}{a_{\sf{B}}\mathcal{R}}\right)\geq0.\label{e239-22}
\end{align}
However, expressions (\ref{e239-2}) and (\ref{e239-22}) are never satisfied; far from the bounce where $a\to\infty$, the second term of (\ref{e239-2}) which always has negative sign dominates and in the limit $a\to a_{\sf{b}}=\mathcal{R}(1-\mathcal{S})$ the expression for {\sf NEC} becomes $-2/\mathcal{R}^{2}(1-\mathcal{S})$ which is also negative because $\mathcal{S}<0$. The same line of reasons can be used to prove the violation of {\sf SEC}. Thus, in model B the {\sf NEC} and {\sf SEC} are never satisfied. Note that, the {\sf NEC} is violated only near the bounce, because expression (\ref{e239-2}) tends to zero as the scale factor gets large values.

By considering the above discussion for energy conditions, we find that again, a phantom scalar field can be used for modeling the behavior of bouncing solution. In the case of model {\sf B} we have
\begin{align}
&\dot{\phi}_{\sf{(eff)B}}^2=-\frac{2 \left(a_{\sf{B}} \mathcal{S}+\left(\mathcal{S}^2-1\right) \mathcal{R}\right)}{a_{\sf{B}}^2 \mathcal{R}},\label{e239-3}\\
&V_{\sf{(eff)B}}=\frac{2 \left(\mathcal{S}^2-1\right)}{a_{\sf{B}}^2}+\frac{5 \mathcal{S}}{a_{\sf{B}} \mathcal{R}}+\frac{3}{\mathcal{R}^2}.\label{e239-4}
\end{align}
Our studies show that the behavior of the above solutions are similar to those of model {\sf A} which in Figure~\ref{figA2} a typical example is demonstrated. By choosing $H_{\sf{B}}$ and $\rho_{\sf{B}}$ as dynamical variables one can rewrite the Friedmann equation as follows
\begin{align}
&\dot{H}=\frac{1-\mathcal{S}^2}{\rho_{0}^2} \rho^{2}+\frac{3}{10}\rho,\label{e239-5}\\
&\dot{\rho}=-H \rho,\label{e239-6}\\
&H^{2}=\frac{1-\mathcal{S}^2}{\rho_{0}^2} \rho^{2}-\frac{3}{5}\rho+\frac{1}{\mathcal{R}^{2}}, \label{e239-7}
\end{align}
where we have dropped the subscript {\sf B}. This system admits two critical points $P_{\sf{B}}^{(\pm)}=(\pm\sqrt{1/\mathcal{R}},0)$ far from the bounce. In these situations we have $\rho\rightarrow0$, hence, the dynamics of equation (\ref{e239-5}) is determined by the second term. Also, near the point of the bounce as specified by $H_{\sf{b}}=0, \rho_{\sf{b}}=3 \rho_{0}^2/10 \mathcal{R}(\mathcal{R}+1)$, we always have $\dot{H}>0$. Therefore, the stability properties of the system is similar to the model {\sf A}.\\

The Friedmann equation (\ref{eom1}) for general function (\ref{e231}) takes the form
\begin{align}\label{e240}
 3H^2=\frac{2 \alpha n(w+1)\Gamma_{\sf{B}}}{(3 w-1) (2 n+3 w-1)}{\sf T}^{\frac{n-2}{w+1}+\frac{3}{2}}+\frac{n (w-1){\sf T}}{(3 w-1) (2 n+w-3)}-\frac{\alpha\Lambda_{\sf{B}}}{2}.
\end{align}
Seeking for a general solution demands that one substitutes the solution (\ref{e232}) into (\ref{e240}) (using the fact that ${\sf T}=(3w-1)\rho$) and solves for the resulting differential equation to find the scale factor. However, the resulting equation cannot be solved analytically for arbitrary values of $w$ and $n$. Nevertheless, for particular values of these parameters a non-singular solution can be obtained as given in expressions (\ref{e234})-(\ref{e238}). But, we are still able to find more general solutions. 
\par
\begin{figure}[h!]
\centering
\centerline{\includegraphics[width=.65\textwidth,trim=10 0 0 0]{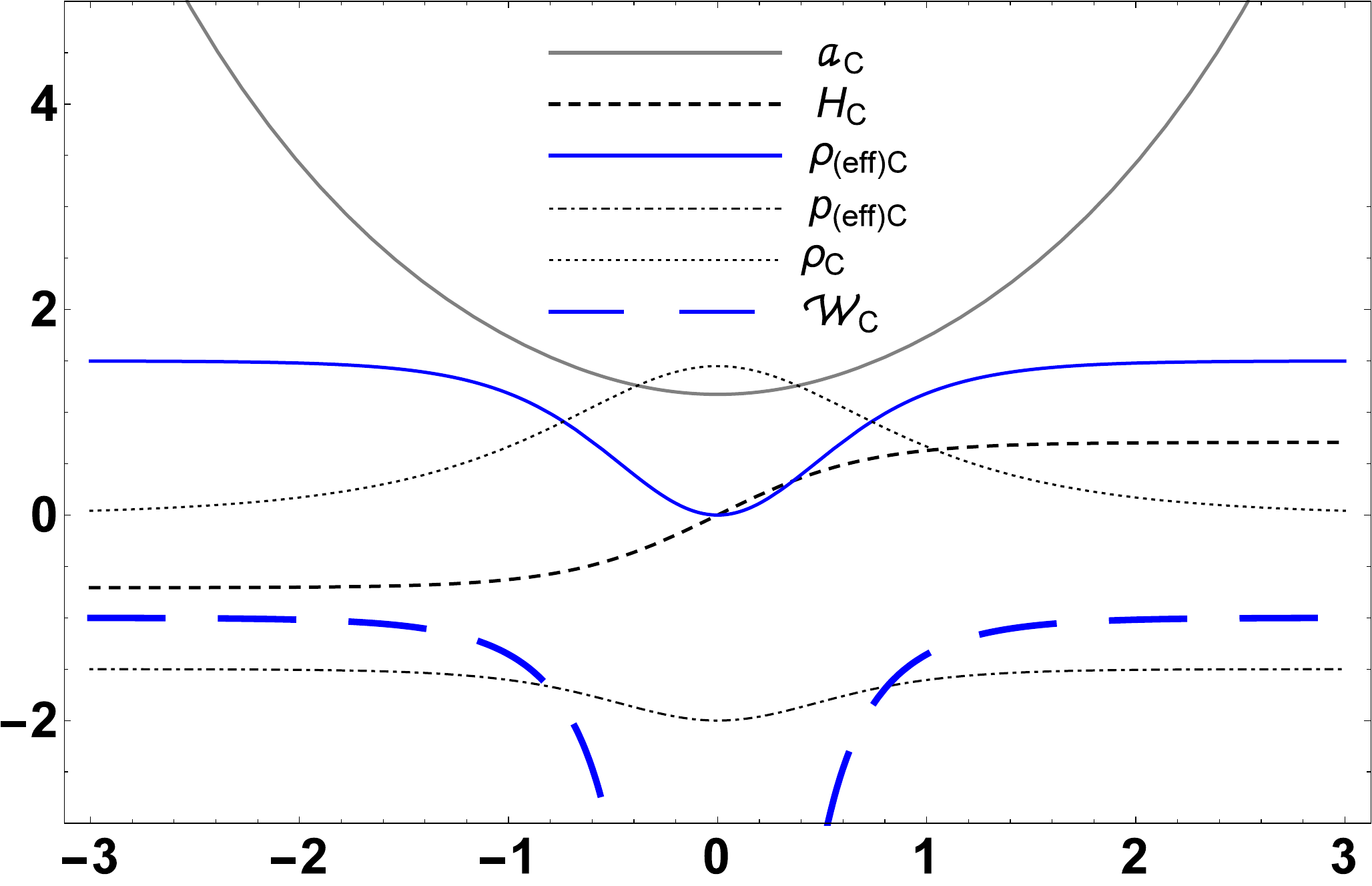}}
\caption{\label{figD}The cosmological quantities related to the bouncing solution of the model {\sf C} for, $n = 3$, $a_0 = 1.3$ $\rho_0 = 2$, $\Lambda_{\sf C}$ = 3, $\alpha=1$ and $Q = 0.5$.}
\end{figure}

As the third type of solutions named as {\sf C}, we work on a bouncing solution for which the governing differential equation is given by
\begin{align}\label{e241}
 3 \left(\frac{a'_{\sf{C}}(t)}{a_{\sf{C}}(t)}\right)^2-\frac{\alpha  \Gamma_{\sf{C}} n\rho_{0}^{\frac{n+1}{2}}}{n+1}a_{\sf{C}}(t)^{\frac{-3(n+1)}{n}}+\frac{\alpha\Lambda_{\sf{C}}}{2}=0,
\end{align}
where we have set $w=1$. Choosing a relation between  $\Gamma_{\sf{C}}$ and $\rho_{0}, a_{0}, n$ and $\Lambda_{\sf{C}}$ as the following
\begin{align}\label{e241-1}
\Gamma_{\sf{C}}=\frac{(n+1)\Lambda_{\sf{C}}\left(\sqrt{2\rho_{0}} a_{0}^{-3/n}\right)^{-n-1}}{n (Q+1)},
\end{align}
equation (\ref{e241}) leads to the following solution for the scale factor
\begin{align}\label{e242}
a_{\sf{C}}(t)=a_{0} \left[(Q+1)\cosh \left(\sqrt{\frac{3\alpha\Lambda_{\sf{C}} }{2}}\frac{(n+1)}{2n}t\pm \frac{\cosh ^{-1}(Q)}{2}\right)\right]^{\frac{2n}{3 n+3}},
\end{align}
where $Q$ is an arbitrary constant and for the Hubble parameter we have
\begin{align}\label{e243}
H_{\sf{C}}(t)=-\sqrt{\frac{\alpha\Lambda_{\sf{C}}}{6}}\tanh \left[\frac{1}{4} \left(\sqrt{6\alpha\Lambda_{\sf{C}}}\frac{ (n+1) }{n}t\pm2\cosh ^{-1}(Q)\right)\right].
\end{align}
By substituting (\ref{e232}) for $w=1$ within the definitions (\ref{re2}), (\ref{re4}) and (\ref{re6}) along with using relation (\ref{e241-1}) we get the effective quantities as follows\footnote{We note that as we are concerned with the solutions that respect the conservation equation (\ref{relation-w}), expression (\ref{e232}) is used for these class of solutions.}
\begin{align}
&\rho_{(\sf{eff})C}(a)=\alpha\Lambda_{\sf{C}}\left[\frac{1}{2}-\frac{\left(\frac{a}{a_0}\right)^{-\frac{3 (n+1)}{n}}}{Q+1}\right],\label{e244}\\
&p_{(\sf{eff})C}(a)=-\alpha\Lambda_{\sf{C}}\left[\frac{\left(\frac{a}{a_0}\right)^{-\frac{3 (n+1)}{n}}}{n(Q+1)}+\frac{1}{2}\right],\label{e245}\\
&\mathcal{W}_{\sf{C}}(a)=\frac{2+n (Q+1) \left(\frac{a}{a_0}\right)^\frac{3 (n+1)}{n}}{n\left[2-(Q+1) \left(\frac{a}{a_0}\right)^\frac{3 (n+1)}{n}\right]}.\label{e246}
\end{align}
We have plotted the cosmological parameters (\ref{e242})-(\ref{e246}) in Figure~\ref{figD}.
Unlike the solution {\sf B}, here, we see that matter creation in the time of bounce leads to a decrease in the effective pressure.
Eliminating the scale factor between expressions (\ref{e244}) and (\ref{e245}) gives the effective {\sf EoS} which can be viewed as the characteristic equation for the model {\sf C}. We therefore get
\begin{align}\label{e2466}
p_{\sf{(eff)}C}= \frac{1}{n}\left(\rho_{(\sf{eff})C}-\frac{\alpha\Lambda_{\sf{C}}}{2}(1+n) \right).
\end{align}
Some of the cosmological properties of model {\sf C} have been investigated in~\cite{Babichev05}. The authors have considered a model of {\sf DE} for which an {\sf EoS} of the form $p_{\sf{DE}}=\gamma(\rho_{\sf{DE}}-\theta)$\footnote{In the original paper they used $\alpha$ and $\rho_{0}$ instead of $\gamma$ and $\theta$, respectively. We changed this character in the present work to prevent ambiguity.} is assumed for a perfect fluid. We therefore observe that if we apply $\rho_{(\sf{eff})}\rightarrow3\rho_{\sf{DE}}$ within equation (\ref{re1}), we obtain the same Friedman equation as the one given in~\cite{Babichev05}. Also, by redefining the parameters as $n\rightarrow1/\gamma$ and $\alpha(1+n)\Lambda_{\sf{C}}/2\rightarrow\theta$ in (\ref{e242}) we will obtain the corresponding solution for the scale factor. These considerations show that the problem of dark fluid with an unusual {\sf EoS} (which may not clearly correspond to a definite Lagrangian) can be explained in the framework of $f({\sf R},{\sf T})$ gravity.
\subsubsection{Type {\sf D} models: solutions which are consistent with the relation $d\rho_{\sf{(eff)}}/d{\sf T}=m$ }\label{sec232}
\begin{figure}[h!]
\centering 
\centerline{\includegraphics[width=.65\textwidth,trim=10 0 0 0]{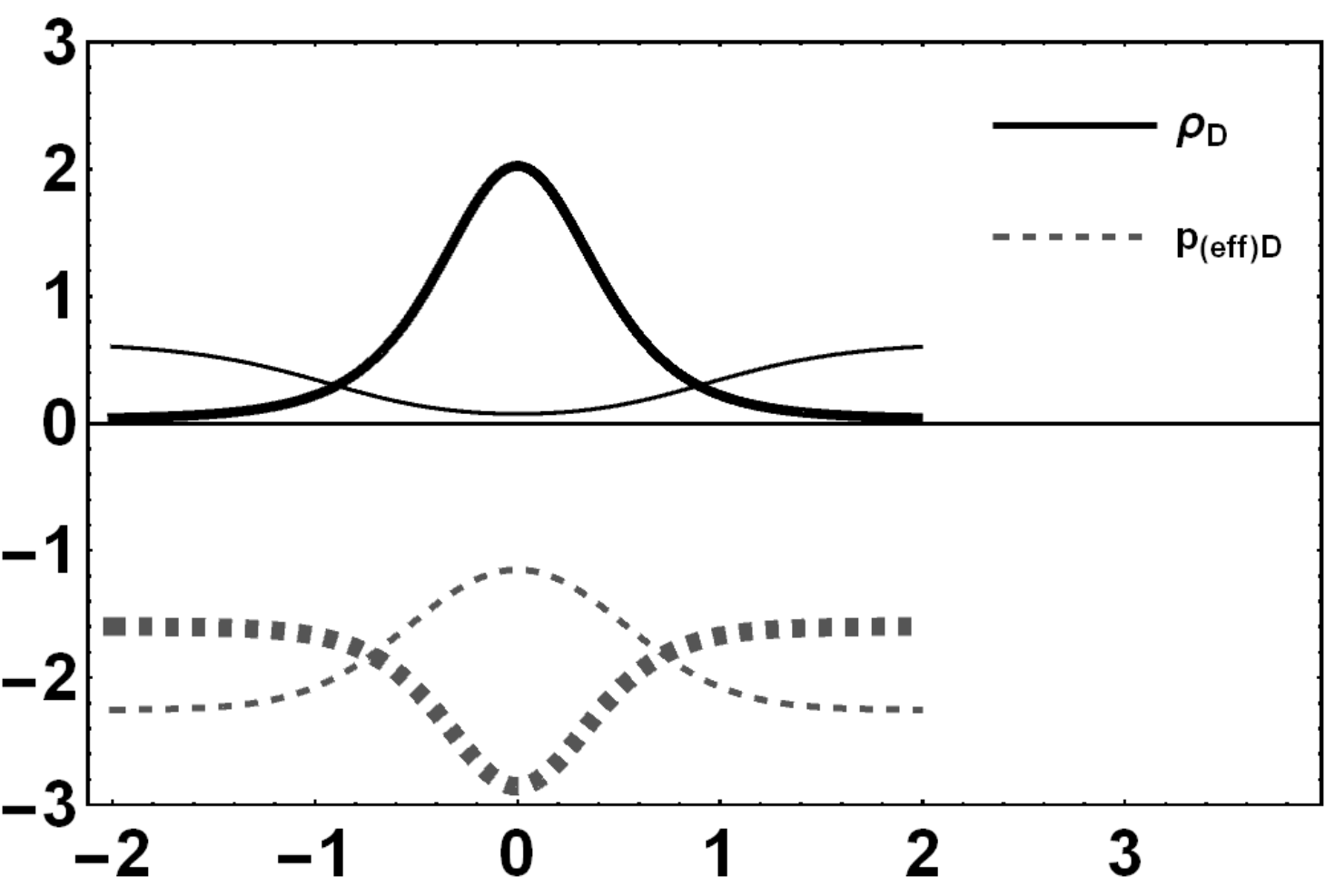}}
\caption{\label{figE}Mutual behaviors of matter energy density and effective pressure related to the bounce solution in model {\sf D}. The thin curves correspond to solution (\ref{e249}) when $\Upsilon=0$. We have set $\alpha=-2.7$, $\Gamma_{\sf D}=-0.1$, $m=-0.8$, $\rho_0=2$, $a_0=1.2$, $\Lambda_{\sf D}=1.2$ for thick curves and $\alpha=-3$, $m=0.2$, $\rho_0=0.1$, $a_0=1.1$, $\Lambda_{\sf D}=-0.2$ for dashed curves.}
\end{figure}

Applying this condition on the definition of effective energy density, i.e., definition  (\ref{re2}), leads to a second order differential equation for $h({\sf T})$ function; the solution then reads
\begin{align}\label{e247}
h_{\sf{D}}({\sf T})=\frac{2\Gamma_{\sf{D}}(w+1)}{3 w-1} {\sf T}^{\frac{3 w-1}{2 (w+1)}}+\frac{2 [m(1-3  w)+1]}{\alpha  (w-3)}{\sf T}+\Lambda_{\sf{D}},
\end{align}
where, $m$ is an arbitrary constant and $\Gamma_{\sf{D}}$, $\Lambda_{\sf{D}}$ are integration constants. Substituting (\ref{e247}) into the conservation equation (\ref{relation-w}), we obtain a first order differential equation for the matter energy density in terms of the scale factor. However, since the mentioned equation cannot be solved for an exact general solution for arbitrary values of $w$ and $m$, we proceed with particular cases. Note that, further investigations may give other exact solutions or even numerical simulations can be utilized to study other solutions. At the present, we work on a particular case $w=1$. The conservation equation (\ref{relation-w}) then yields
\begin{align}\label{e248}
\rho_{\sf{D}}(a)=\frac{\left(\zeta -\alpha\Gamma_{\sf{D}}a_{\sf{D}}^{3}\right)^2}{8 a_{\sf{D}}^6 m^2},
\end{align}
where we have again used ${\sf T}=(3w-1)\rho$. In the limit $a\rightarrow\infty$ one obtains $\Gamma_{\sf{D}}^{2}\alpha^{2}/8m^{2}$ from solution (\ref{e248}). As can be seen, in the model {\sf D} the matter energy density evolves from a non-vanishing initial value far from the bounce event. This property cannot be seen in the previous models. To obtain the solution for the scale factor, we substitute (\ref{e247}) in the Friedmann equation (\ref{re1}) for $w=1$, which gives
\begin{align}\label{e249}
a_{\sf{D}}(t)=\left[\frac{\Gamma_{\sf{D}} \zeta }{\omega }-\frac{\Delta}{\omega} \cosh \left(\sqrt{\frac{3 \alpha\omega }{4 m}}t\right)- \Upsilon\sinh \left(\sqrt{\frac{3 \alpha\omega }{4 m}}t\right)\right]^{1/3},
\end{align}
where 
\begin{align}
&\zeta=\alpha\Gamma_{\sf{D}}\pm2 m \sqrt{2\rho_{0}},~~~~~~~~~~~~~~~~~~~~~~~~\omega=\alpha  {\Gamma_{\sf{D}}}^{2}-2m\Lambda_{\sf{D}},\label{e250}\\
&\Upsilon=\sqrt{a_{0}^{6}+\frac{\zeta }{\alpha  \omega }\left(\zeta -2 \alpha a_{0}^{3} \Gamma_{\sf{D}}\right)},~~~~~~~~~~
\Delta=\Gamma_{\sf{D}} \zeta -\omega  a_{0}^{3}.\label{e2501}
\end{align}
From solution (\ref{e249}), we can obtain the time at which the bounce occurs as well as the radius of the Universe at the moment of bounce, as
\begin{align}\label{e251}
t^{\sf{(b)}}_{\sf{D}}=\left(\frac{3 \alpha\omega }{m}\right)^{-1/2}\log \left(\frac{\Delta-\omega \Upsilon}{\Delta+\omega \Upsilon }\right),
\end{align}
and
\begin{align}\label{e252}
a^{\sf{(b)}}_{\sf{D}}=\left[{\frac{\Gamma_{\sf{D}} \zeta +\sqrt{\Delta^{2}-\omega^{2}\Upsilon^{2}}}{\omega }}\right]^{1/3}.
\end{align}
Differentiating solution (\ref{e249}) with respect to time gives the Hubble parameter as follows
\begin{align}\label{e253}
H_{\sf{D}}=\sqrt{\frac{\alpha  \omega }{12m}}\frac{\omega \Upsilon\cosh \left(\sqrt{\frac{3 \alpha\omega }{4 m}}t\right)+\Delta\sinh \left(\sqrt{\frac{3 \alpha\omega }{4 m}}t\right)}{\Delta\cosh \left(\sqrt{\frac{3 \alpha\omega }{4 m}}t\right)+\omega \Upsilon\sinh \left(\sqrt{\frac{3 \alpha\omega }{4 m}}t\right)-\Gamma_{\sf{D}}\zeta},
\end{align}
and from definitions of the effective quantities we find
\begin{align}
&\rho_{\sf{(eff)D}}=\frac{\left(\zeta -\alpha\Gamma_{\sf{D}}a_{\sf{D}}^3\right)^2}{4m a_{\sf{D}}^6 }-\frac{\alpha\Lambda_{\sf{D}}}{2},\label{e253d0}\\
&p_{\sf{(eff)D}}=\frac{1}{4} \left[2 \alpha \Gamma_{\sf{D}} \sqrt{\frac{\left(\zeta-\alpha  \Gamma_{\sf{D}}a_{\sf{D}}^3 \right)^2}{m^2a_{\sf{D}}^6}}+\frac{\left(\zeta-\alpha  \Gamma_{\sf{D}}a_{\sf{D}}^3\right)^2}{m a_{\sf{D}}^6}+2 \alpha  \Lambda_{\sf{D}}\right],\label{e253d1}\\
&\mathcal{W}_{\sf{D}}=\frac{\alpha\left(2 \Lambda_{\sf{D}}  m-\alpha  \Gamma_{\sf{D}}^2\right) a_{\sf{D}}^6+\zeta ^2}{\left(\zeta -a_{\sf{D}}^3 \alpha  \Gamma_{\sf{D}} \right)^2-2m\alpha  \Lambda_{\sf{D}}a_{\sf{D}}^6}.\label{e256}
\end{align}
By inspection of expressions (\ref{e253d0}) and (\ref{e253d1}), we observe that model {\sf D} corresponds to the following {\sf EoS}
\begin{align}\label{e2566}
p_{\sf{(eff)D}}=\rho_{\sf{(eff)D}}-\frac{\alpha \Gamma_{\sf{D}}}{\sqrt{m}}\sqrt{\rho_{\sf{(eff)D}}+\frac{\alpha \Lambda_{\sf{D}}}{2}}+\alpha\Lambda_{\sf{D}}.
\end{align}
In view of what we discussed in the paragraph right after equation (\ref{e239-1}), model {\sf D} corresponds to a generalized {\sf EoS} with $\beta=1$. This may be interesting since the most bouncing models have been obtained for $\beta=-1$. Note that, the behavior of obtained cosmological quantities for two values of $\zeta$ as given in (\ref{e250}), are similar to those of model {\sf C}. By choosing the model parameters so that the term including $sinh$ in $(\ref{e249})$ disappears, we arrive at a different model. In this case, the behavior of cosmological quantities is the same as the case for which $\Upsilon\neq0$. However, the evolution of matter energy density and the effective pressure are different. We typically plot these quantities for both situations in Figure~\ref{figE}. The thick curves belong to the solution (\ref{e249}) and the thin ones show the solution with $\Upsilon=0$. The energy condition considerations show that models of type {\sf D} lead to the violation of {\sf NEC} near the bounce event. Also the phase space and the scalar field representation of this model are the same as model {\sf A}.
\section{Matter bounce solutions in $f({\sf R},{\sf T})$ gravity}\label{Sec4}
\begin{figure}[h!]
\centering 
\includegraphics[width=.48\textwidth,trim=10 0 2 0]{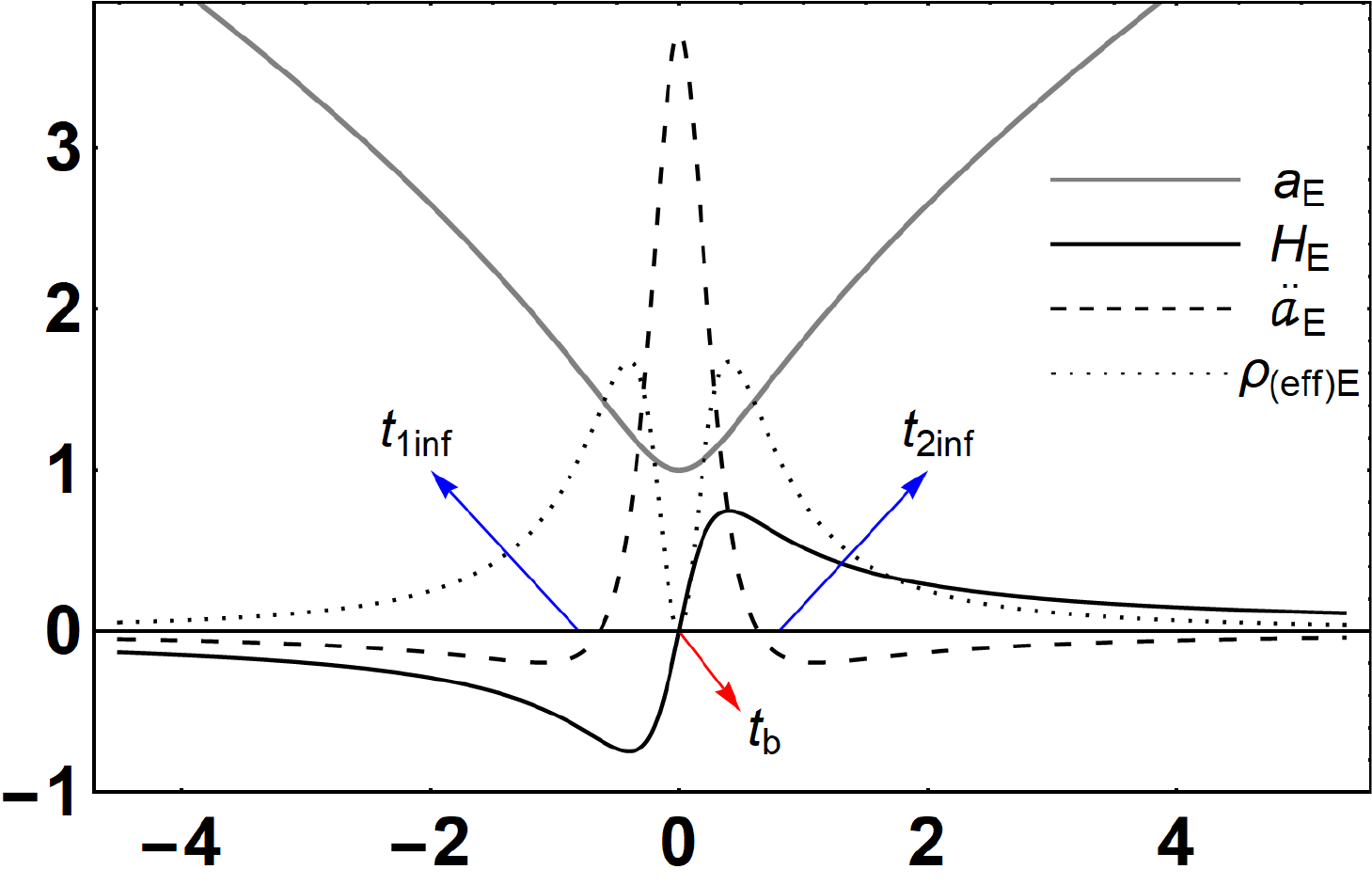}
\hfill
\includegraphics[width=.48\textwidth,origin=c,angle=360]{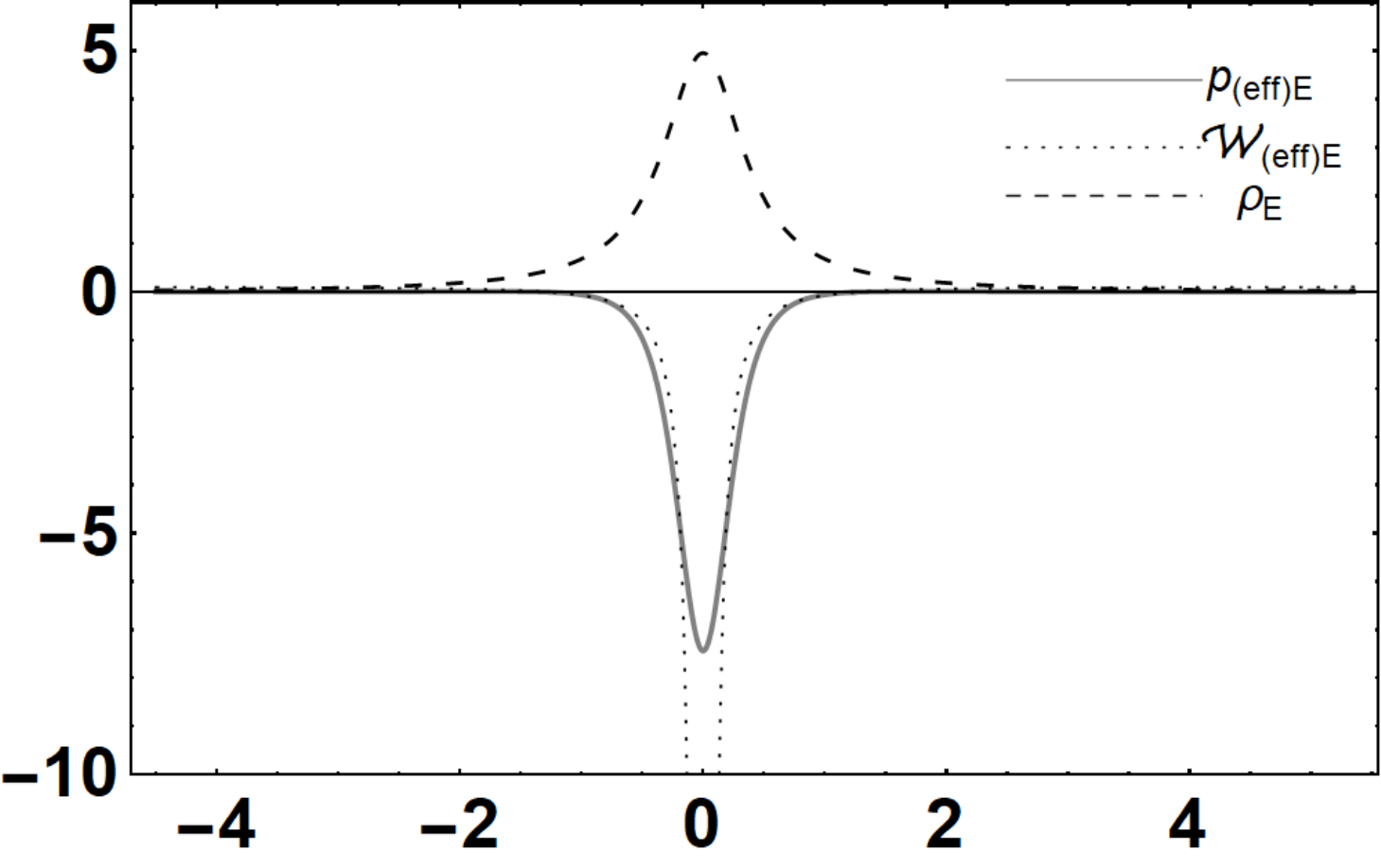}
\caption{\label{figF1} The behavior of scale factor, Hubble parameter, acceleration of the Universe and effective energy density in matter bounce scenario. Right panel: The behavior of effective pressure, effective {\sf EoS} and matter energy density for model {\sf E}. The model parameters have been set as, $\mathbf{M}=0.6$, $\alpha\Gamma_{\sf E}=-0.011$ and $\mathbf{Z}=1$.}
\end{figure}

In this section, we deal with the well-known matter bounce scenario which can be established through the models that obey (\ref{e231}) and (\ref{e232}). We specify these class of solutions as type {\sf E} models. A branch of the matter bounce scenarios have been discussed with the characteristic scale factor
\begin{align}\label{e41}
a_{\sf{E}}(t)=\left(\mathbf{Q} t^2+\mathbf{Z}\right)^\mathbf{M},
\end{align}
where $\mathbf{Q}$, $\mathbf{Z}$ and $\mathbf{M}$ are positive constants. Note that $\mathbf{Q}=u\rho_{\sf{max}}$, with $u=2/3,~3/4,~4/3$, $\mathbf{Z}=1$ and $\mathbf{M}=1/3$~\cite{Cai11,Odintsov14,Odintsov151} and  $\mathbf{M}=1/4$~\cite{Cai15} has been used in the literature. The scale factor (\ref{e41}) gives the following expression for the Hubble parameter
\begin{align}\label{e42}
H_{\sf{E}}(t)=\frac{2 \mathbf{M}\mathbf{Q} t}{\mathbf{Q} t^2+\mathbf{Z}}.
\end{align}
Note that in model {\sf E} the Hubble parameter and its time derivatives never diverge since all of them are proportional to negative powers of $\mathbf{Q} t^2+\mathbf{Z}$. By eliminating the time parameter between (\ref{e41}) and (\ref{e42}), we get the Hubble parameter in terms of the scale factor and thus the Friedmann equation can be obtained as
\begin{align}\label{e43}
3H_{\sf{E}}^{2}(a)=12 \mathbf{Q} \mathbf{M}^2 \left[a_{\sf{E}}^{-\frac{1}{\mathbf{M}}}- \mathbf{Z} a_{\sf{E}}^{-\frac{2}{\mathbf{M}}}\right].
\end{align}
From another side, substituting (\ref{e231}) into (\ref{eom1}) together with using (\ref{e232}), the Friedmann equation in terms of the scale factor reads
\begin{align}\label{e44}
3H^{2}(a)= \frac{2 \alpha \Gamma_{\sf{E}} (w+1)n \rho_0^{\frac{n-2}{w+1}+\frac{3}{2}} \left(3 w-1 \right)^{\frac{n-2}{w+1}+\frac{1}{2}}}{2 n+3 w-1}a^{-\frac{3 (1-2n-3w)}{2n}}+\frac{(w-1)n \rho _0}{2 n+w-3}a^{-\frac{3 (w+1)}{n}},
\end{align}
where we have used subscript {\sf E} for integration constant $\Gamma_{\sf E}$ and we have set $\Lambda_{\sf{E}}=0$. Comparing equations (\ref{e44}) and (\ref{e43}) we find two different type of solutions which only one of them can be accepted. A consistent solution is valid for
\begin{align}
&n=\frac{12 \mathbf{M}}{6\mathbf{M}-1},~~~~~~~~~~~~~~~~~~~~~~w= \frac{4}{6 \mathbf{M}-1}-1,\label{e45}\\
&\mathbf{Z}=\frac{16 \Gamma_{\sf{E}}\alpha \rho_{0} (2-3 \mathbf{M})}{12 \mathbf{M} (3 \mathbf{M}-2)+3},~~~~~~~~\mathbf{Q}=\frac{3(1-2\mathbf{M}) \rho_{0}}{4 \mathbf{M} (6 \mathbf{M}-1)}.\label{e46}
\end{align}
Eliminating $\rho_{0}$ between $\mathbf{Z}$ and $\mathbf{Q}$ leads to
\begin{align}\label{e47}
\mathbf{Q}=\frac{9 (2 \mathbf{M}-1)^2}{64 \alpha\Gamma_{\sf{E}}\ \mathbf{M} (3 \mathbf{M}-2)} \mathbf{Z}.
\end{align}
Valid solutions for which $\{\mathbf{Q},\mathbf{Z},\mathbf{M}\}>0$ holds, are given  by
\begin{align}
&\frac{1}{2}<\mathbf{M}<\frac{2}{3}~~~~\mbox{which corresponds to}~~~\frac{1}{3}<w<1~~\mbox{for}~~~\alpha\Gamma_{\sf{E}}<0,\label{e48}\\
&\mathbf{M}>\frac{2}{3}~~~~~~~~~~\mbox{which corresponds to}~~~-1<w<\frac{1}{3}~~\mbox{for}~~~\alpha\Gamma_{\sf{E}}>0.\label{e49}
\end{align}
As can be seen,  in the context of $f({\sf R},{\sf T})$ gravity, there can be found a matter bounce solution for every values of $w$ (except for $w=1/3,1$). The value $w=-1$ can be accessed for large values of $\mathbf{M}$ (see relations (\ref{e45})).
From (\ref{e232}) and (\ref{e45}) we see that for models of type {\sf E} we have $\rho_{\sf{E}}=\rho_{0}a_{\sf{E}}^{-1/\mathbf{M}}$. The effective quantities are then obtained as
\begin{align}
&\rho_{\sf{(eff)E}}=\frac{12 \mathbf{M}^{2} \mathbf{Q}}{\mathbf{Z}}a_{\sf{E}}^{-\frac{2}{\mathbf{M}}} \left(a_{\sf{E}}^{\frac{1}{\mathbf{M}}}-1\right)=\frac{12 \mathbf{M}^2 \mathbf{Q}^2 t^2}{\mathbf{Z} \left(\mathbf{Q} t^2+1\right)^2},\label{e50}\\
&p_{\sf{(eff)E}}=-\frac{4\mathbf{M}\mathbf{Q}}{\mathbf{Z}}a_{\sf{E}}^{-\frac{2}{\mathbf{M}}}\left((3\mathbf{M}-1) a_{\sf{E}}^{\frac{1}{\mathbf{M}}}-3\mathbf{M}+2\right)\nonumber\\
&=-\frac{4\mathbf{M}\mathbf{Q}}{\mathbf{Z}}\frac{(3\mathbf{M}-1) \mathbf{Q} t^2+1}{\left(\mathbf{Q} t^2+1\right)^2}\label{e51},\\
&\mathcal{W}_{\sf{E}}=\frac{1}{3} \left(\frac{1}{\mathbf{M}(1- a^{\frac{1}{\mathbf{M}}})}+\frac{1}{\mathbf{M}}-3\right)=-1+\frac{1}{3\mathbf{M}}\left(1-\frac{1}{\mathbf{Q} t^2}\right).\label{e52}
\end{align}
Model {\sf E} corresponds to an effective {\sf EoS} as
\begin{align}\label{e53}
p_{\sf{(eff)E}}^{\pm}=\left(\frac{2}{3 \mathbf{M}}-1\right)\rho_{\sf{(eff)E}}\pm\frac{2 \mathbf{Q}}{\mathbf{Z}}\sqrt{\mathbf{M}^2-\frac{\mathbf{Z}}{3 \mathbf{Q}}\rho_{\sf{(eff)E}}}-\frac{2 \mathbf{M} \mathbf{Q}}{\mathbf{Z}}.
\end{align}
Estimating effective pressure (\ref{e51}) in the limiting times $t\to 0$ and $t\to\pm\infty$, indicates that the effective {\sf EoS} follows the expression $p_{\sf{(eff)E}}^{-}$ far from the bounce and obeys $p_{\sf{(eff)E}}^{+}$ near the bounce. In $t\to 0$  pressure (\ref{e51}) gives $p_{\sf{(eff)E}}=-4\mathbf{M}\mathbf{Q}/\mathbf{Z}$ which is consistent with $p_{\sf{(eff)E}}^{-}$ and in $t\to\pm\infty$ we have  $p_{\sf{(eff)E}}=0$ which can be explained only with $p_{\sf{(eff)E}}^{+}$. Figure~\ref{figF1} shows the behavior of different quantities. As is seen in the left panel, the scale factor decreases till reaching a minimum non-zero value at $t=t_{{\sf b}}$ where the Hubble parameter vanishes. The Universe experiences four phases during its evolution from pre-bounce to post bounce. Before the bounce occurs, the Universe has been within an accelerated contracting regime till the first inflection point ($t_{\sf 1inf}<t_{\sf b}$) is reached at which the accelerations vanishes. At this point the Hubble parameter maximizes in negative direction and correspondingly the effective energy density gains a peak value. The Universe then enters a decelerating contracting regime so that its velocity decreases in negative direction. The collapse of Universe halts at the bounce time after which the Universe goes into an accelerating expanding phase where both $\ddot{a}_{\sf E}>0$ and $H_{\sf E}>0$. Once the second inflection point ($t_{\sf 2inf}>t_{\sf b}$) is reached, the Universe enters a decelerating expanding regime so that the speed of expansion decreases at later times.\par
\begin{figure}[h!]
\centering 
\includegraphics[width=.48\textwidth,trim=10 0 2 0]{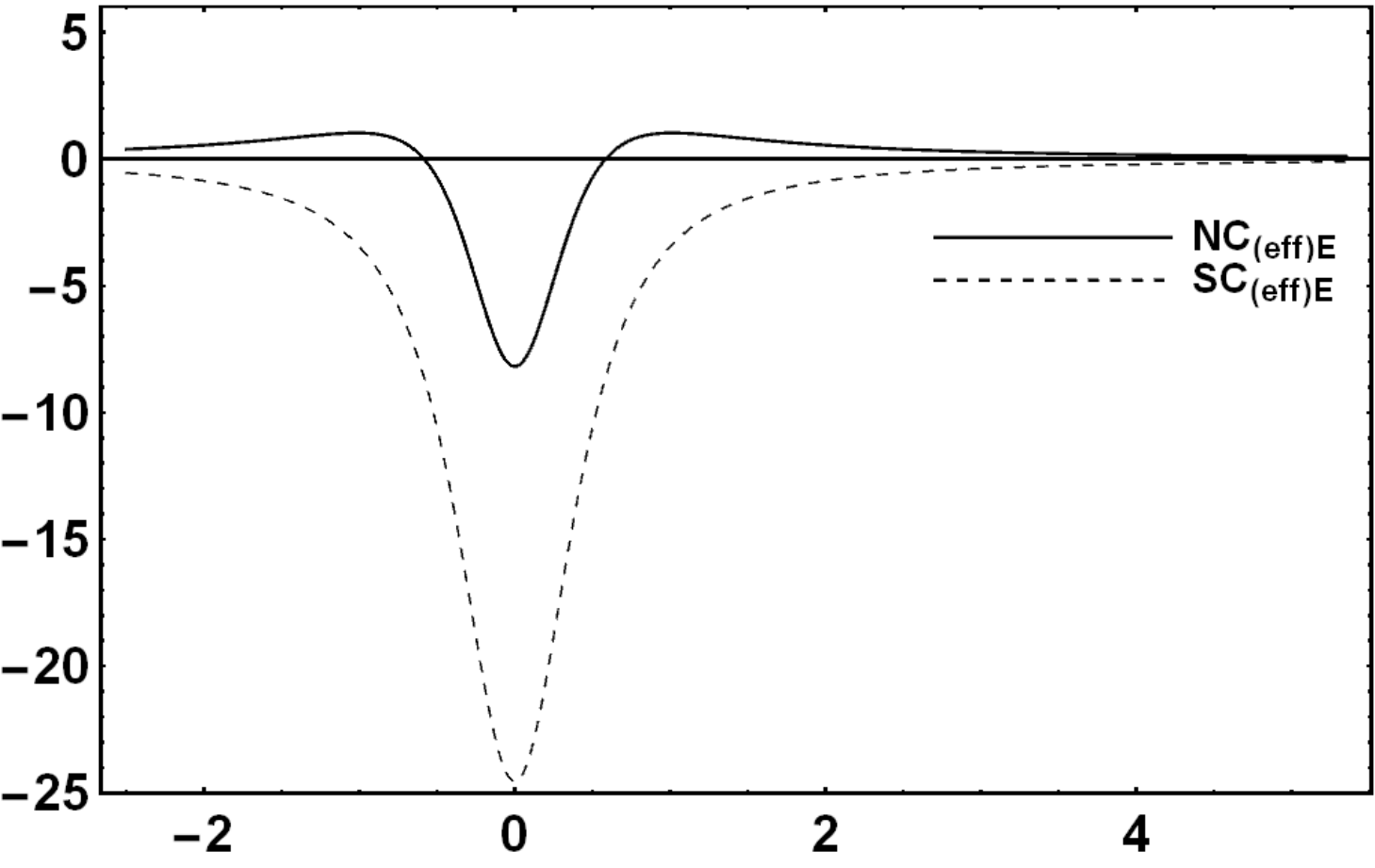}
\hfill
\includegraphics[width=.48\textwidth,origin=c,angle=360]{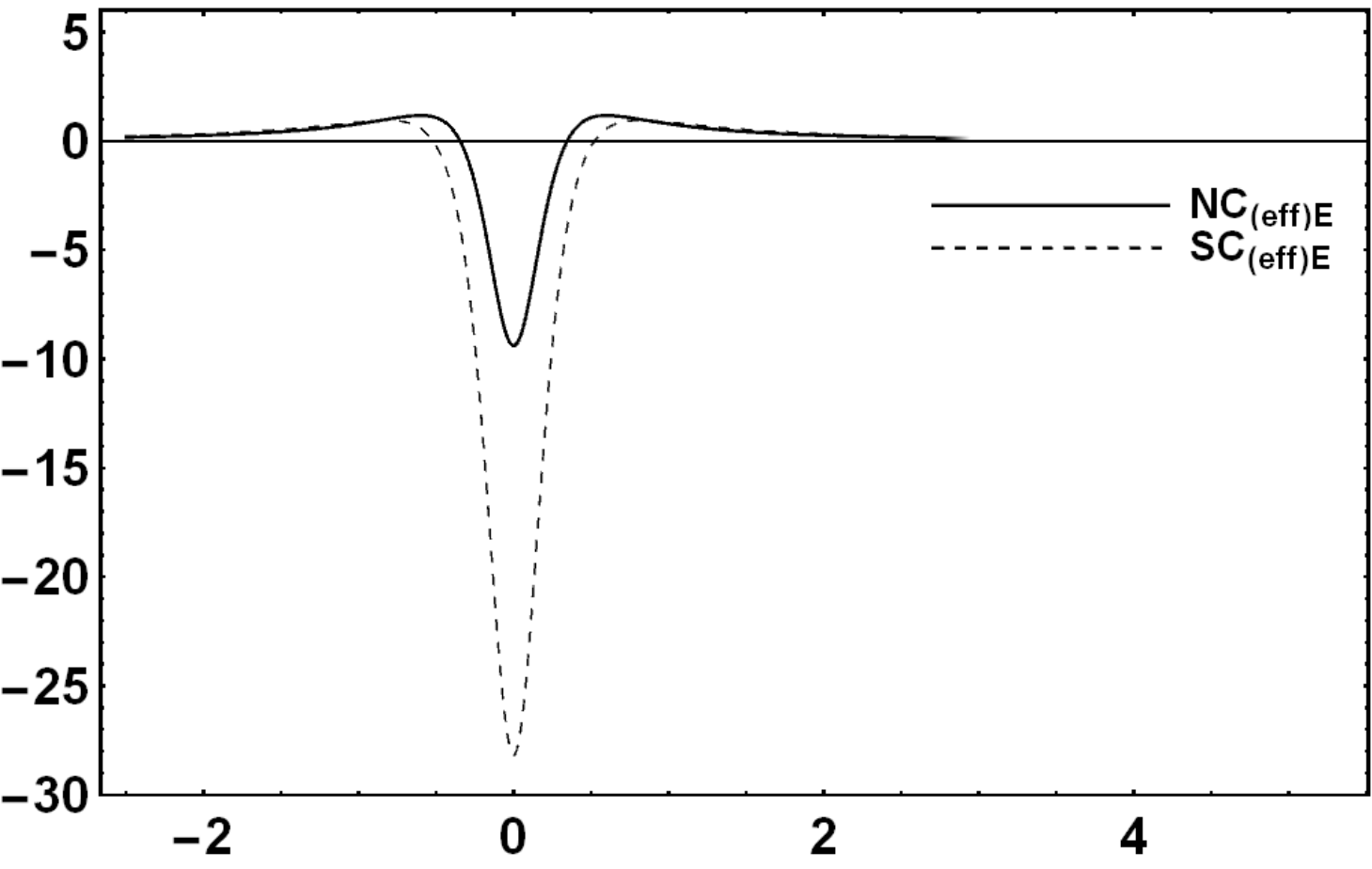}
\caption{\label{figF2} The behavior of the {\sf NEC} and {\sf SEC} for the model {\sf E}, left panel for $\mathbf{M}=0.7$ and right panel for $\mathbf{M}=0.3$. We see that These conditions are violated near the bounce and it is possible to be respected far away from the bounce (see equations (\ref{e54}) and (\ref{e55}) and the subsequent discussion).}
\end{figure}
\begin{figure}[h!]
\centering 
\includegraphics[width=.55\textwidth,trim=10 0 -5 0]{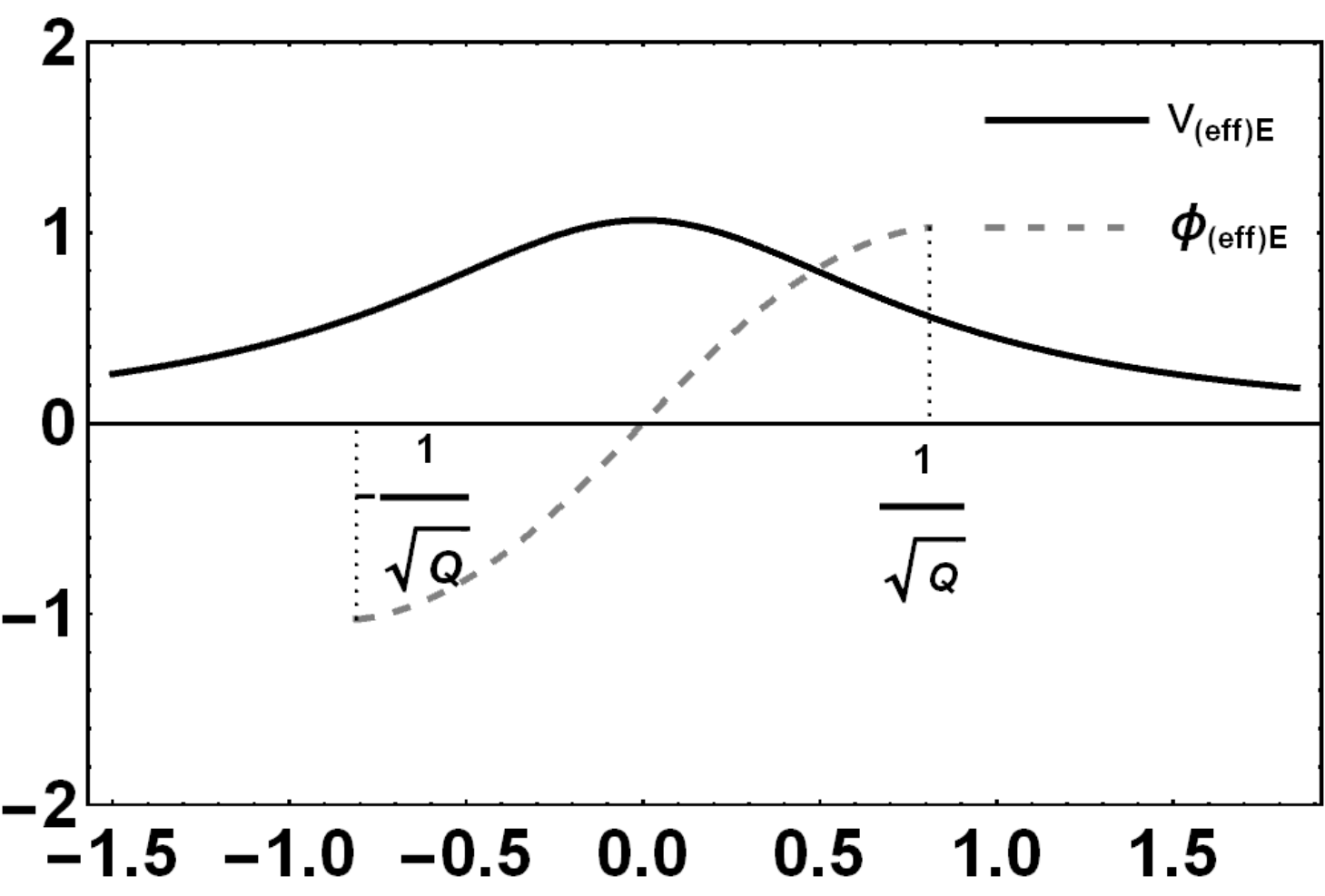}
\caption{\label{figF3} The effective phantom scalar field and the corresponding potential for the model {\sf E}. As before, the horizontal axis shows the time values. In the model {\sf E}, the {\sf NEC} is violated only near the bounce (within the period  $-1/\sqrt{\mathbf{Q}}<t<1/\sqrt{\mathbf{Q}}$) in contrast to the other discussed models. A phantom scalar field representation can be used to equivalently describe the model which can be valid when the {\sf NEC} is violated.}
\end{figure}

Let us now check the behavior of  {\sf NEC}$_{\sf{(eff)}}$ and ${\sf SEC}_{\sf{(eff)}}$ which for model {\sf E} take the following form as
\begin{align}
&{\sf NEC}_{\sf{(eff)E}}=p_{\sf{(eff)E}}+\rho_{\sf{(eff)E}}=\frac{4 \mathbf{M}\mathbf{Q} \left(\mathbf{Q} t^2-1\right)}{\mathbf{Z} \left(\mathbf{Q} t^2+1\right)^2},\label{e54}\\
&{\sf SEC}_{\sf{(eff)E}}=p_{\sf{(eff)E}}+3\rho_{\sf{(eff)E}}=\frac{12 \mathbf{M}\mathbf{Q} \left[(1-2\mathbf{M})\mathbf{Q} t^2-1\right]}{\mathbf{Z} \left(\mathbf{Q} t^2+1\right)^2}.\label{e55}
\end{align}
It is obvious that in the expression (\ref{e54}), the sign of $\mathbf{Q} t^2-1$ determines the validity of {\sf NEC}. We therefore find that near the bounce, {\sf NEC} is violated within the range $-1/\sqrt{\mathbf{Q}}<t<1/\sqrt{\mathbf{Q}}$, and out of this range it is preserved. The {\sf SEC} is violated within a larger time interval, i.e., in $-1/\sqrt{\mathbf{Q}(1-2\mathbf{M})}<t<1/\sqrt{\mathbf{Q}(1-2\mathbf{M})}$ for $0<\mathbf{M}<1/2$ and is always violated for $\mathbf{M}>1/2$. In figure~\ref{figF2} we have plotted the expression of NEC and SEC for two different values of $\mathbf{M}$, i.e., for $\mathbf{M}=0.3$ and $\mathbf{M}=0.7$. We see again that, in the background of $f({\sf R},{\sf T})$ gravity a bouncing behavior corresponds to violating the {\sf NEC}.
We then may conclude that scalar representation type {\sf E } models can be constructed by a phantom field within the time interval where {\sf NEC} is violated (notice that from solution (\ref{e52}) we see that for the same intervals in which the {\sf NEC} is violated, we have $\mathcal{W}_{\sf{E}}<-1$). In this cases, the kinetic energy of scalar field is negative. Thus, for a phantom scalar field in model {\sf E} we obtain
\begin{align}
&\phi_{\sf{(eff)E}}=-\frac{8}{3} \sqrt{\frac{\mathbf{M}}{\mathbf{Z}}} \left[\arcsin\left(\sqrt{\mathbf{Q}} t\right)-\sqrt{2} \arctan\left(\sqrt{\frac{2\mathbf{Q}t}{1-\mathbf{Q} t^2}}\right)\right],\label{e56}\\
&V_{\sf{(eff)E}}=\frac{2 \mathbf{M}\mathbf{Q} \left[(6\mathbf{M}-1)\mathbf{Q} t^2+1\right]}{\mathbf{Z} \left(\mathbf{Q} t^2+1\right)^2}.\label{e57}
\end{align}
In Figure~\ref{figF3} we have presented typical diagrams for the effective phantom field and its corresponding potential.
To see the stability properties of model {\sf E}, we rewrite the field equations as follows
\begin{align}
&2\dot{H}_{\sf E}=3\frac{2\mathbf{M}-1}{6\mathbf{M}-1}\rho\left(1-\frac{2\rho}{\rho_{0}}\right),\label{e58}\\
&\dot{\rho}_{\sf E}=-\frac{H}{\mathbf{M}} \rho,\label{e59}\\
&3H_{\sf E}^{2}=-9\mathbf{M}\frac{2\mathbf{M}-1}{6\mathbf{M}-1}\rho\left(1-\frac{\rho}{\rho_{0}}\right)\label{e60}.
\end{align}
The analysis of these equations is similar to those of the previous models. Firstly, note that only for $1/6<\mathbf{M}<1/2$, equation of (\ref{e60}) leads to the standard Friedmann equation when the correction term is absent. The system (\ref{e58}) and (\ref{e59}) have two fixed points with coordinates ${\sf P}_{\sf{E}}^{(\pm)}=(\pm0,0)$ which correspond to the limit $\rho\to0$. The bounce event occurs at $\rho_{\sf E}=\rho_{0}$ at which we have $\dot{H}_{\sf E}=-3(2\mathbf{M}-1)/6\mathbf{M}-1>0$. The stability of the system can be analyzed similar to the way used in the previous sections. A simple study shows the stability properties are analogous to the other models; the evolution of the Universe begins from an unstable state, then passing through an unstable bounce phase and finally reaches a stable state. 

\begin{sidewaystable}
\begin{center}
\caption{Different bouncing solutions and their main properties\textsuperscript{*}}
\begin{tabular}{l @{\hskip 0.05in} l@{\hskip 0.2in} l@{\hskip 0.15in} l@{\hskip 0.15in} l@{\hskip 0.15in} l@{\hskip 0.15in}}\hline\hline

Models& \thead[c]{$a(t)$}&$p_{\sf{(eff)}}$& $h({\sf T})$&$\rho$&$w$\\[0.5 ex]
\hline\\

{\sf A}&\makecell[l]{$\frac{1}{\mathfrak{A}}\left[\cosh \left(\sqrt{-\frac{\mathcal{P}}{3}t}\right)-\sinh\left(\sqrt{-\frac{\mathcal{P}}{3}t}\right)\right]\times$\\$ \left\{\mathfrak{B}+\frac{(w^2-1)\mathcal{P}\rho_{0}}{2(3w-1)}\left[\sinh \left(\sqrt{-3\mathcal{P}} t\right)+\cosh \left(\sqrt{-3\mathcal{P}} t\right)-1\right]\right\}^{2/3}$}&$\mathcal{P}=cons.$&$\frac{2}{\alpha }\left(\mathcal{P}+\frac{w}{1-3 w}{\sf T}\right)$&$\rho_{0}a^{-3}$&$>1/3$\\[1.5 ex]

{\sf B}&$\mathcal{R}\left[\cosh\left(\frac{t}{\mathcal{R}}\right)-\mathcal{S}\right]$&\makecell[l]{$-\frac{\rho}{3}+\frac{2 \mathcal{S} \sqrt{3\mathcal{R}^2 \left(\mathcal{S}^2-1\right)\rho+9}}{3\mathcal{R}^2 \left(\mathcal{S}^2-1\right) }$\\+$\frac{2}{\mathcal{R}^2\left(\mathcal{S}^2-1\right)}$}& $\frac{\Gamma {\sf T}^{2}}{2}-\frac{7}{4\alpha}{\sf T}+\Lambda$&$\rho_{0}a^{-1}$&$-1/5$\\[1.5 ex]

{\sf C}&$a_{0} \left[(Q+1)\cosh \left(\sqrt{\frac{3\alpha\Lambda}{2}}\frac{(n+1)}{2n}t\pm \frac{\cosh ^{-1}(Q)}{2}\right)\right]^{2n/3(n+1)}$&$\frac{1}{n}\left[\rho-\frac{\alpha\Lambda_{\sf{C}}}{2}(1+n) \right]$&$\frac{2\Gamma {\sf T}^{\frac{n+1}{2}}}{n+1}-\frac{{\sf T}}{\alpha}+\Lambda$&$\rho_{0}a^{-6/n}$&1\\[1.5 ex]

{\sf D}&$\left[\frac{\Gamma \zeta }{\omega }-\frac{\Delta}{\omega} \cosh \left(\sqrt{\frac{3 \alpha\omega }{4 m}}t\right)- \Upsilon\sinh \left(\sqrt{\frac{3 \alpha\omega }{4 m}}t\right)\right]^{1/3}$&$\rho-\frac{\alpha \Gamma}{\sqrt{m}}\sqrt{\rho+\frac{\alpha \Lambda}{2}}+\alpha\Lambda$&$2\Gamma \sqrt{{\sf T}}-\frac{(2m-1){\sf T}}{\alpha}+\Lambda$&$\frac{\left(\zeta -\alpha\Gamma a^{3}\right)^2}{8 a^6 m^2}$&1\\[1.5 ex]

{\sf E}&$\left(\mathbf{Q} t^2+\mathbf{Z}\right)^\mathbf{M}$&\makecell[l]{$\left(\frac{2}{3 \mathbf{M}}-1\right)\rho$\\$\pm\frac{2 \mathbf{Q}}{\mathbf{Z}}\sqrt{\mathbf{M}^2-\frac{\mathbf{Z}}{3 \mathbf{Q}}\rho}-\frac{2 \mathbf{M} \mathbf{Q}}{\mathbf{Z}}$}&\makecell[l]{$\frac{2 \Gamma (w+1)}{2 n+3 w-1} {\sf T}^{\frac{n-2}{w+1}+\frac{3}{2}}$\\$-\frac{2 (n-1) }{\alpha  (2 n+w-3)}{\sf T}+\Lambda$}&$\rho_{0}a^{-3 (w+1)/n}$&$w$\\
\hline\hline
\multicolumn{6}{l}{\footnotesize * The subscripts {\sf A,..,E} and \lq\lq{}{\sf eff}\rq\rq{} are dropped for abbreviation.}
\end{tabular}
\label{Tab1}
\end{center}
\end{sidewaystable}
%
\section{Stability of the bouncing models}\label{comm}
\begin{figure}[ht]
\centering
\centerline{\includegraphics[width=1.2\textwidth,trim=10 0 0 0]{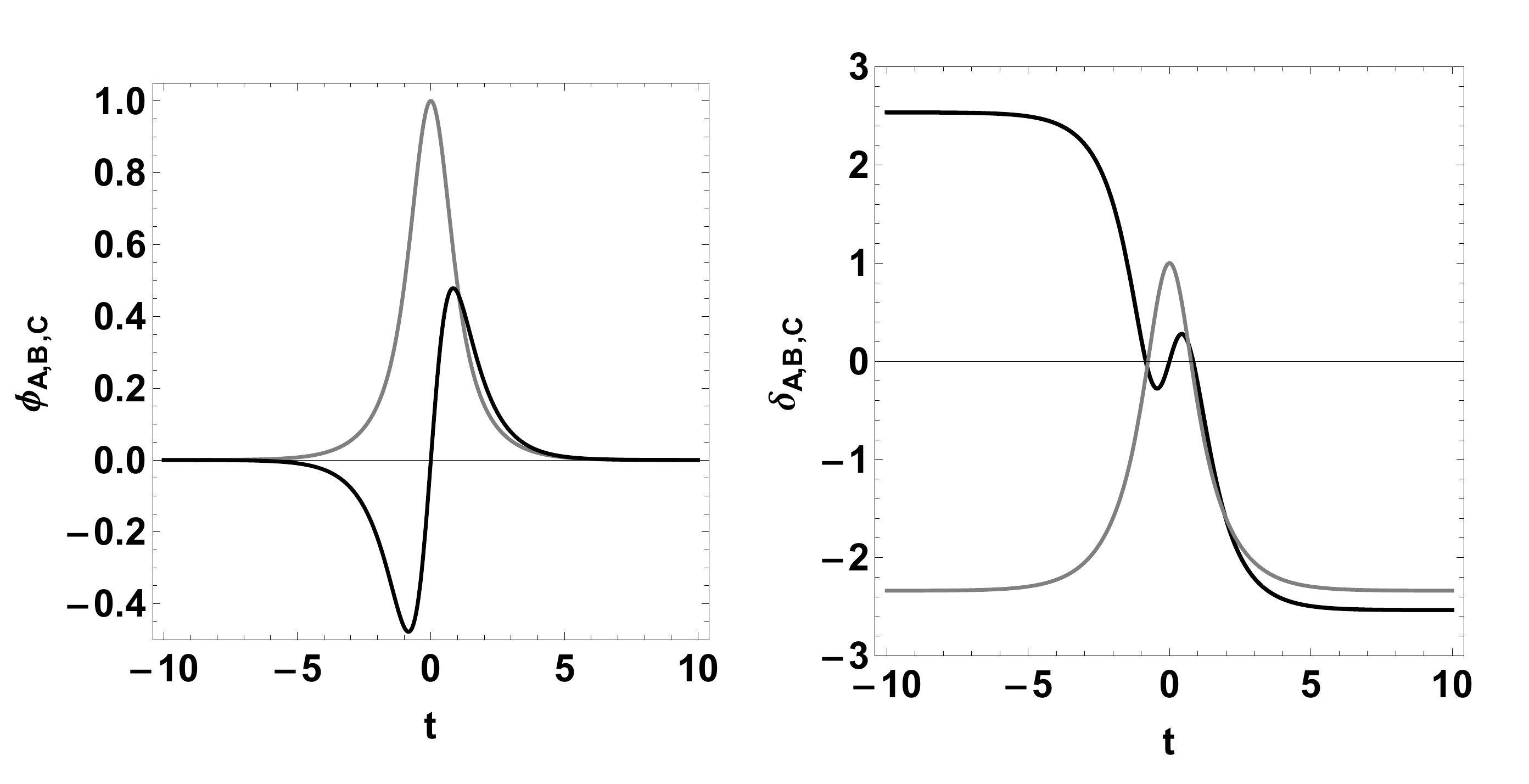}}
\caption{\label{dpertABC} Evolution of $\delta$ and $\Psi$ for models {\sf A}, {\sf B} and {\sf C}. The behavior of these quantities are similar for the first three model. The black curves are plotted for initial values $\delta(0)=0, \delta'(0)=1, \Phi(0)=0$ and $\Phi'(0)=1$ and the gray ones are drawn for $\delta(0)=1, \delta'(0)=0, \Phi(0)=1$ and $\Phi'(0)=0$. }
\end{figure}

In this section we verify the wholesomeness of the bouncing models which has been introduced in the previous sections through considering the possibility of occurring serious instabilities. We therefore examine the evolution of scalar-type perturbations in the discussed models within the metric formalism. Since $f(\sf{R,T})$ gravity introduces unusual coupling of matter to curvature part of its action, the evolution of matter density perturbations (specially, as the effect of bounce event on the evolution of matter perturbations is not obvious) can be problematic. In order to study such type of perturbations we consider the matter density perturbations in $f({\sf R},{\sf T})={\sf R}+\alpha\kappa^{2} h({\sf T})$ models for a fat {\sf FLRW} metric in the longitudinal gauge 
\begin{align}\label{sec52-1}
ds^2=-(1+2\Phi)dt^{2}+a(t)^{2}(1-2\Psi)\delta_{ij}dx^{i}dx^{j},
\end{align}
where the metric scalar perturbations $\Phi$ and $\Psi$ are functions of four coordinates $(t,x,y,z)$, generally. In the current work we shall obtain necessary equations for models including a barotropic perfect fluid with equation of state $p=w\rho$ and a general $h({\sf T})$ function. In this respect, the authors of~\cite{Alvarenga13} have already considered the matter perturbations in a narrow class of $f({\sf R},{\sf T})$ models\footnote{Paper~\cite{Alvarenga13} has considered models in which the conservation of {\sf EMT} is respected. These modes accept $h({\sf T})=\sqrt{{\sf T}}$~\cite{Shabani172}.} for a pressure-less perfect fluid. The perturbations of {\sf EMT} in the longitudinal gauge are given by~\cite{Tsujikawa08}
\begin{align}\label{sec52-2}
&\delta {\sf T}^{t}_{\,t}=-\delta \rho_{m},\\
&\delta {\sf T}^{i}_{\,t}=\frac{1}{a}(1+w)\rho v_{,i},\\
&\delta {\sf T}^{t}_{\,i}=-a (1+w)\rho v_{,i},\\
&\delta {\sf T}^{i}_{\,j}=w\delta_{ij} \delta \rho,
\end{align}
where, $v$ is a covariant velocity perturbation~\cite{Malik05}. Using the background equations (\ref{eom1}) and (\ref{eom2}) we obtain the following equations for the scalar perturbations in Fourier space
\begin{align}\label{sec52-3}
2\frac{k^2}{a^2}\Psi+6H(H\Phi+\dot{\Psi})=\delta \Sigma_{\,t}^{t}+\frac{1}{2}\mathcal{F}\delta \sf{T},
\end{align}
as the {\sf ADM} energy constraint ($\mathcal{G}_{\,t}^{t}$ component of the field equation, if we rewrite (\ref{fRT field equations}) as $\mathcal{G}_{\,\mu}^{\nu}=\Sigma_{\,\mu}^{\nu}$),
\begin{align}\label{sec52-4}
H\Phi+\dot{\Psi}=-\frac{1}{3}\int{\delta \Sigma_{\,i}^{t}dx^{i}},
\end{align}
as the {\sf ADM} momentum constraint ($\mathcal{G}_{\,i}^{t}$ component). Moreover, we have
\begin{align}\label{sec52-5}
\Phi-\Psi=0,
\end{align}
as the {\sf ADM} propagation equation ($\mathcal{G}_{\,i}^{j}-1/3\delta_{\,i}^{j}\mathcal{G}_{\,l}^{l}$ component),
\begin{align}\label{sec52-6}
&6\left[\ddot{\Psi}+2H\dot{\Psi}+H\dot{\Phi}+2(H^{2}+\dot{H})\Phi\right]\nonumber\\
&-2\frac{k^2}{a^2}\Phi=\delta \Sigma_{\,i}^{i}-\delta \Sigma_{\,t}^{t}+\mathcal{F} \delta \sf{T},
\end{align}
as the perturbed version of Raychaudhuri equation ($\mathcal{G}_{\,i}^{i}-\mathcal{G}_{\,t}^{t}$ component),
\begin{align}\label{sec52-7}
\delta \sf{R}=-\left(\delta \Sigma+2\mathcal{F} \delta \sf{T}\right),
\end{align}
as the trace equation ($\mathcal{G}_{\,\mu}^{\mu}=\Sigma_{\,\mu}^{\mu}$),
\begin{align}\label{sec52-8}
\dot{\delta}+\eta H\delta+\xi\left(\frac{k^2}{a^2}v-3\dot{\Psi}\right)=0,
\end{align}
as the time component of perturbed {\sf EMT} conservation and finally
\begin{align}\label{sec52-9}
\dot{v}+3H\sigma v-\Phi+\lambda\delta=0,
\end{align}
as the spatial component of perturbed {\sf EMT} conservation\footnote{For more details and also the utilized terminology, see~\cite{Hwang01}.}. Note that equations (\ref{sec52-3})-(\ref{sec52-9}) are the most general equations describing scalar perturbations in minimal $f({\sf R},{\sf T})$ gravity for condition $F=1$ when a barotropic perfect fluid is included. These equations are not independent, so that it is possible to obtain one equation from another one; for example (\ref{sec52-7}) follows from multiplying (\ref{sec52-3}) by $2$ then adding it to (\ref{sec52-6}) and using (\ref{sec52-12}). In the above equations and relations we have used the following definitions for the source terms, which appear in the right hand sides of field equation (\ref{fRT field equations}), its trace and in equation (\ref{relation}) when is written as $\nabla_{\beta}\sf{T}_{\,\alpha}^{\beta}=\Sigma_{\alpha}$, respectively
\begin{align}
&\Sigma_{\,\mu}^{\nu}=\left(\kappa^{2}+\mathcal{F}\right){\sf T}_{\,\mu}^{\nu}-w\rho \mathcal{F} g_{\,\mu}^{\nu},\label{sec52-101}\\
&\Sigma=\left(\kappa^{2}+\mathcal{F}\right){\sf T}-4w\rho\mathcal{F},\label{sec52-102}\\
&\Sigma_{\,\alpha}=\frac{1}{\kappa^{2}+\mathcal{F}}\left[w\nabla_{\alpha}\left(\rho\mathcal{F}\right)-\frac{1}{2}\mathcal{F}\nabla_{\,\alpha}{\sf T}- \nabla_{\beta}\mathcal{F} \sf{T}_{\,\alpha}^{\beta}\right]. \label{sec52-103}
\end{align}
The gauge-invariant density contrast in the longitudinal gauge is defined as
\begin{align}\label{sec52-11}
\delta=\frac{\bar{\delta}\rho}{\rho}+3Hv
\end{align}
and the perturbed Ricci scalar for the {\sf FRLW} metric can be obtained as
\begin{align}\label{sec52-12}
\bar{\delta}{\sf R}=-2\left[3\ddot{\Psi}+12 H\dot{\Psi}+3H\dot{\Phi}+6\dot{H}\Phi+12H^{2}\Phi-\frac{k^{2}}{a^{2}}\left(\Phi-2\Psi\right)\right].
\end{align}
Also, using definitions in (\ref{sec52-101})-(\ref{sec52-103}), the coefficients in (\ref{sec52-8}) and (\ref{sec52-9}) can be obtained as follows 
\begin{align}\label{sec52-13}
&\eta=-3(1+w)(3w-1)\frac{N_\eta}{D_\eta}H\rho,~~~~~\sigma=3\frac{N_\sigma}{D_\sigma},\\
&\xi=(1+w)\left[\frac{\kappa^{2}+\frac{1}{2}(3-w)\mathcal{F}+(1+w)(3w-1)\rho\mathcal{F}'}{\kappa^{2}+\mathcal{F}}\right]^{-1},\\
&\lambda=\frac{1}{w+1}\left[\frac{1}{2}(1-w)\frac{\mathcal{F}}{\kappa^{2}+\mathcal{F}}-w\right],
\end{align}
where
\begin{align}\label{sec52-13w}
&N_\eta=-(1+w)\mathcal{F}\mathcal{F}'+(1+w)(3w-1)\rho\mathcal{F}'^{2}\nonumber\\
&-\frac{\kappa^{2}(w+3)}{2}\mathcal{F}'-(1+w)(3w-1)(\kappa^{2}+\mathcal{F})\rho\mathcal{F}'',\nonumber\\
&D_\eta=\left[\kappa^{2}+\frac{1}{2}(3-w)\mathcal{F}+(1+w)(3w-1)\rho\mathcal{F}'\right]^{2},
\end{align}
and
\begin{align}\label{sec52-15}
&\nonumber\\
&N_\sigma=\frac{1}{2}(1-w)\mathcal{F}-w(\kappa^{2}+\mathcal{F})+(3w^{2}+2w-2)\rho\mathcal{F}'\nonumber\\
&D_\sigma=\kappa^{2}+\frac{1}{2}(3-w)\mathcal{F}+(1+w)(3w-1)\rho\mathcal{F}',
\end{align}
and prime denotes derivative with respect to the trace. Note that wherever is needed we have used $\dot{\rho}$ which can be obtained from the {\sf EMT} conservation equation (\ref{relation}). Now, using equations  (\ref{sec52-8}) and (\ref{sec52-9}) we arrive at the evolutionary equation for the matter perturbation, as follows,
\begin{align}\label{sec52-17}
\ddot{\delta}+\mathcal{D}_{1}\dot{\delta}+\mathcal{D}_{2}\delta+\xi\left[-3\ddot{\Phi}-3(3\sigma+2)H\dot{\Phi}+\frac{k^{2}}{a^{2}}\Phi\right]=0.
\end{align}
From equations (\ref{sec52-7}) and (\ref{sec52-12}) we obtain a dynamical equation for the perturbed potential $\Phi$, as
\begin{align}\label{sec52-18}
2\left[3\ddot{\Phi}+15H\dot{\Phi}+(6\dot{H}+12 H^{2}+\frac{k^{2}}{a^{2}})\Phi\right]+\theta\delta=0,
\end{align}
where we have used solution (\ref{sec52-5}) and defined the following coefficients
\begin{align}\label{sec52-19}
&\mathcal{D}_{1}=\left[(2+3\sigma)H+\eta-\frac{\dot{\xi}}{\xi}\right],\\
&\mathcal{D}_{2}=\left(\dot{\eta}-\frac{\dot{\xi}}{\xi}\eta+3\sigma\eta H-\lambda\xi\frac{k^{2}}{a^{2}}+2\eta H\right),\\
&\theta=\Bigg[\kappa^{2}+(3-5w)\mathcal{F}-(1+w)(3w-1)\mathcal{F}'\rho\Bigg]\rho.
\end{align}
Hence, we have two differential equations (\ref{sec52-17}) and (\ref{sec52-18}) along with relation (\ref{sec52-3}) to be solved for $\delta$ and $\Psi$. Obviously, the coefficients of these equations are some complicated functions of model properties, $H, a, \rho, {\sf T}, \mathcal{F}, \mathcal{F}', \mathcal{F}''$, and thus it may not be possible to obtain exact solutions. However, one can resort to numerical methods or even in the case of stiff equations, obtain approximated solutions. We have plotted the evolution of $\delta$ and $\Phi$ in Figure~\ref{dpertABC} for models {\sf A}, {\sf B} and {\sf C}. Numerical simulations show that the behavior of perturbations are typically similar in {\sf A}, {\sf B} and {\sf C} models for the same initial values. We have sketched two set of plots in Figure~\ref{dpertABC}. These models show a zero value for the fluctuations when the initial values $\delta(0)=0, \delta'(0)=1, \Phi(0)=0$ and $\Phi'(0)=1$ are assumed. In this case, the fluctuations tend to zero (left panel) and constant values (right panel) in the regime of large times, see the black curves in Figure~\ref{dpertABC}. As another case, the fluctuations increase from zero to a maximum finite value in the period of bounce if the initial data are set as $ \delta(0)=1, \delta'(0)=0, \Phi(0)=1$ and $\Phi'(0)=0$, see the gray curves. As can be seen from the evolution of scalar perturbations across the bounce, though small temporary fluctuations, no instability occurs during the bounce, nor does it happen in the limit of large times for these models. 

\begin{figure}[ht]
\centering
\centerline{\includegraphics[width=1.2\textwidth,trim=10 0 0 0]{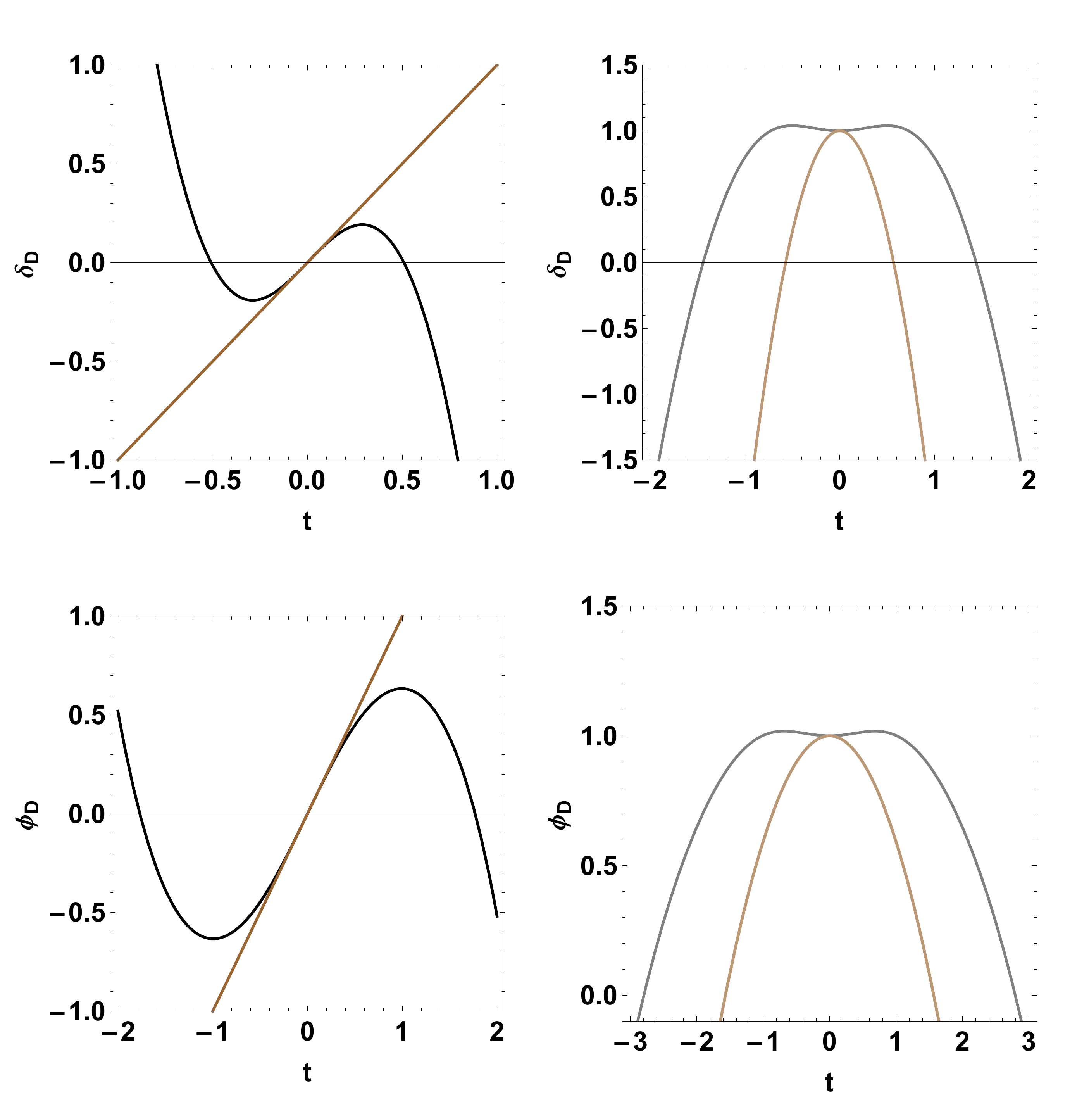}}
\caption{\label{dpertD} The evolution of $\delta$ and $\Psi$ for model {\sf D}. The Left panels are plotted for initial values $\delta(0)=0, \delta'(0)=1, \Phi(0)=0$ and $\Phi'(0)=0$. The right panels are drawn for $\delta(0)=1, \delta'(0)=0, \Phi(0)=1$  and $\Phi'(0)=0$. The black curves are the solutions of equations approximated for large times and the brown ones are solutions in the limit of small times approximated around the bounce time.}
\end{figure}

Unfortunately, the system of differential equations (\ref{sec52-17}) and (\ref{sec52-18}) become stiff for models {\sf D} and {\sf E} so as it is not possible to plot reasonable diagrams for $\delta$ and $\Phi$. In this case, we proceed to obtain approximate solutions. For models {\sf E} and {\sf D} (in case in which $\Upsilon=0$), equations (\ref{sec52-17}) and (\ref{sec52-18}) at times near the bounce will take the following forms
\begin{align}
&\ddot{\delta}+\frac{1}{2} \left(2 \mathcal{D}_{2}^{0}+\xi^{0}\theta^{0}\right) \delta +2\left(3H'^{0}+\frac{k^2}{{a^{0}}^2}\right)\xi^{0} \Phi=0,\label{sec52-201}\\
&6 \ddot{\Phi}+\theta^{0} \delta+ \left(\frac{2k^2}{{a^{0}}^2}+12 H'^{0}\right)\Phi=0,\label{sec52-202}
\end{align}
for which the solutions up to second order can be found as
\begin{align}
&\delta=\delta_{i}+\delta'_{i}t+\left[-\frac{1}{4} \left(2 \mathcal{D}_{2}^{0}+\xi^{0}\theta^{0}\right)\delta_{i}-\left(3H'^{0}+\frac{k^2}{{a^{0}}^2}\right)\xi^{0} \Phi_{i}\right]t^{2},\label{sec52-211}\\
&\Phi=\Phi_{i}+\Phi'_{i}t+\left[-\frac{1}{12}\theta^{0}\delta_{i}- H'^{0}\Phi_{i}-\frac{k^{2}}{6{a^{0}}^{2}}\Phi^{i}\right]t^{2},\label{sec52-212}
\end{align}
where the superscript ``0" denotes the values of quantities in the limit $t\to0$ and the subscript $i$ shows the initial values required for integrations. As we see, the solutions are stable in the period of bounce. Note that other coefficients except those which are shown in equations (\ref{sec52-201}) and (\ref{sec52-202}) vanish in the limit $t\to0$. In the limit of large times, we have numerically plotted the evolution of the matter contrast $\delta$ and the potential $\Phi$ in Figure~\ref{dpertD}. As can be seen, from (\ref{sec52-211}) and (\ref{sec52-212}), it is obvious that there happens no instability at the period of bounce in models {\sf D} and {\sf E}, however, far away from the bounce point, both $\delta$ and $\Phi$ increase dramatically in model {\sf D}, see Figure~\ref{dpertD}.
 For model {\sf E}, in the limit of large times, $t\to\infty$, we get
\begin{align}\label{sec52-22}
&\ddot{\delta}-3\xi^{\infty}\ddot{\Phi}=0,\\
&\ddot{\Phi}=0,
\end{align}
with the solutions
\begin{align}\label{sec52-21}
&\delta=\delta_{i}+\delta'_{i}t,\\
&\Phi=\Phi_{i}+\Phi'_{i}t.
\end{align}
Therefore, depending on the initial values, the perturbations in model {\sf E} can grow before and after the bounce. Therefore we conclude that, though the scalar-type perturbations in models {\sf D} and {\sf E} behave regularly at the bounce point, these solutions are unstable asymptotically.

\section{Singular solutions}\label{singularsols}
In this section we seek for possible solutions that exhibit singular behavior, specially the big-bang singularity.
Some singular solutions have been already studied in the literature which we suffice to give a short discussion for them. Presuming the {\sf EMT} conservation (i.e., supposing ${\sf T}={\sf T}_{0} a^{-3(1+w)}$), equation (\ref{relation-w}) can be solved to give the following solution
\begin{align}\label{sec6-1}
h({\sf T})=2{\sf Q}_{1}\frac{w+1}{3 w+1} {\sf T}^{\frac{3}{2}-\frac{1}{w+1}}+{\sf Q}_{2},
\end{align}
where $\sf{Q}_{1}$ and $\sf{Q}_{2}$ are some constants. In~\cite{Shabani172} the authors have analytically shown that  for the case of pressure-less perfect fluid, i.e., for $w=0$, one finds
\begin{align}\label{sec6-2}
a_{\sf{SIN}}^{\sf{I}}(t)=\left(\frac{3 }{256}\right)^{1/3}\left[\sqrt{\Omega_{0}^{\rm{(p)}}}H_0 t \left(8 \sqrt{3}-3 \beta t\right)\right]^{2/3},
\end{align}
which shows a singular behavior. This singular solution is the only one which respects the {\sf EMT} conservation. Relaxing such a constraining condition, one can obtain other kind of solutions. Note that function (\ref{sec6-1}) is the only form that respects the {\sf EMT} conservation, that is, to obtain another solution one should assume some suitable form for $h(\sf{T})$. For example, in~\cite{Shabani172} for $f({\sf R},{\sf T})={\sf R}+\alpha \kappa^2 {\sf T}$ the authors have obtained 
\begin{align}
&\rho(a)=\rho_{0} a^{\frac{6 (\alpha-1)}{2-3\alpha}},\label{sec6-3}\\
&a_{\sf{SIN}}^{\sf{II}} (t)=\left(\frac{3}{2}\right)^{\frac{2-3 \alpha }{6(1- \alpha)}} \left[\frac{3(1-\alpha ) \sqrt{\Omega_{0}} H_{0}t}{\sqrt{2-3 \alpha}}\right]^{\frac{2-3\alpha}{3 (1-\alpha)}},\label{sec6-4}
\end{align}
for a pressure-less matter. In this case singular solution can be found for some valid range of values of the model constant $\alpha$.
Also, the following solution has been obtained for models of form $f({\sf R},{\sf T})={\sf R}+\alpha \kappa^2 {\sf T}^{-1/2}$
\begin{align}
&\rho (a)=2^{-2/3}\left(\frac{\alpha+2{\rho_{0}}^{3/2}}{a^{9/2}}-\alpha\right)^{2/3}\label{sec6-5}\\
&a_{\sf{SIN}}^{\sf{III}}(t)=\left(\frac{27}{256}\right)^{1/9}\left[\alpha+2(3H_{0}^{2}\Omega_{0})^{3/2}\right]^{2/9}t^{2/3}.\label{sec6-6}
\end{align}
 As a new class of (big-bang) singular models, we examine the models which admit the following solution
\begin{align}\label{sec6-7}
a_{\sf{SIN}}^{\sf{IV}}(t)=a_{0}\left(\frac{t}{t_{0}}\right)^{\ell},
\end{align} 
where $\ell$ being a real positive number and $t_{0}$ being the time at which the scale factor gets the present value $a_{0}$. Any solution has to satisfy two of the three equations (\ref{eom1})-(\ref{relation-w}), for a presumed scale factor function. In this case, two unknown functions ${\sf T}(a)$ and $h(\sf{T})$ should be obtained via solving the resulted equations. To proceed further we make use of the following anzats
\begin{align}\label{sec6-8}
h(\sf{T})=\sf{C_1} \sf{T}^{\mu }+\sf{C_2} \sf{T},
\end{align} 
where constants ${\sf C}_i$'s and $\mu$ should be fixed according to the following considerations. Substituting (\ref{sec6-7}) and (\ref{sec6-8})
in (\ref{relation-w}) and solving for ${\sf T}(a)$ we get
\begin{align}\label{sec6-9}
{\sf T}(a)=\left\{\frac{\alpha  {\sf C_1} \mu  (w-3)\sf{T}_0}{ \Big[\alpha  {\sf C_1} \mu  (w-3) {\sf T}_0^{\mu }+(w-1){\sf T}_0 \Big]\left(\frac{a}{a_0}\right)^{-\frac{6 (\mu -1) (w+1)}{2 \mu  (w+1)-3 w+1}}+{\sf T}_0 (1-w)}\right\}^{\frac{1}{1-\mu }},
\end{align} 
where an integration constant has been set so that ${\sf T}(a=a_{0})={\sf T_0}$. To ensure that three functions  (\ref{sec6-7}), (\ref{sec6-8}) and  (\ref{sec6-9}) provide an analytic solution, they must satisfy at least one of the equations  (\ref{eom1}) and (\ref{eom2}). Substituting these functions in (\ref{eom1}) shows that we have a solution only for the case of stiff fluid, i.e., for $w=1$. Therefore, a singular solution can be obtained provided that
\begin{align}\label{sec6-10}
{\sf C_1} =-\frac{2\ell (3\ell-2) {\sf T}_0^{\frac{1}{3 \ell-2}}}{t_{0}^{2}\alpha  \kappa^{2} },~~{\sf C_2}=-\frac{1}{\alpha },~~\mu =\frac{1}{2-3\ell},~~\frac{1}{3}<\ell<\frac{2}{3},~~w=1.
\end{align} 
Thus, for conditions (\ref{sec6-10}) we obtain
\begin{align}\label{sec6-11}
&{\sf T}={\sf T}_{0} \left(\frac{a}{a_{0}}\right)^{6-\frac{4}{\ell}},\\
&f(\sf{R},\sf{T})=\sf{R}-\kappa^{2}\sf{T}+2 \ell (3 \ell-2) t_{0}^{-2}\left(\frac{{\sf T}}{{\sf T}_{0}}\right)^{\frac{1}{2-3 \ell}}.
\end{align} 
Therefore, besides the non-singular solutions obtained in the present paper, one can still find a set of singular solutions. In this brief section in addition to addressing some previous results we obtained a new singular model, though a coherent study can be performed in order to deal with possible conditions for which big-bang singularity would occur; however working on this issue is beyond the scope of the present paper and comprehensive studies on this subject will be reported elsewhere. It is worthwhile to mention that some studies have been already made to consider other forms of singular solutions, e.g., \cite{Houndjo13}. Beside the above results, Bianchi type I cosmological model with magnetized strange quark matter in the framework of $f ({\sf R}, {\sf T})$ gravity have been investigated and it is found that the model begins with big-bang and ends with big rip~\cite{sahoo}. Using Lie point symmetry analysis method, the authors of~\cite{alisymm} have shown that for a Bianchi type I spacetime, both singular (big-bang) and nonsingular solutions could exist subject to the type of specified symmetry.

Recently, the authors of \cite{Awad18}, have considered some cosmological features of $f(\mathcal{T})$ gravity (where here $\mathcal{T}$ denotes torsion scalar) using the dynamical system approach both generally and for some specific forms of $f(\mathcal{T})$ functions. The core of their studies is taking the advantage of this fact that the torsion scalar can be used interchangeably with the Hubble parameter (i.e., $\mathcal{T}=-6H^{2}$). Thus, the field equations reduce to a single equation (in the case of pressure-less matter) in the form of $\dot{H}=\mathcal{F}(H)$, since the matter density can also be rewritten as a function of the Hubble parameter. Briefly, they have shown that in $f(\mathcal{T})$ gravity a single equation (which can be interpreted as a simple one dimensional dynamical system) can govern the dynamics of field equations. Benefiting this useful result they investigated phase space portraits of various cosmological evolutions such as, singular and non-singular solutions. Likewise, one may be motivated to utilize such an approach in order to investigate the cosmological solutions of $f({\sf R}, {\sf T})$ gravity (especially, in the case of present work, i.e., the function given in (\ref{minimal})) through phase portrait diagrams. However, looking at equations (\ref{eom2}) and (\ref{relation-w}) one finds that it is impossible to obtain an equation like $\dot{H}=\mathcal{F}(H)$ so that it reflects full information of the field equations. In this case for an assumed function $h(\sf{T}), $ we have a two dimensional dynamical system without any further reduction. Thus, the procedure proposed in~\cite{Awad18} would be generally failed in $f(\sf{\sf R}, {\sf T})$ gravity.
\section{Concluding remarks}\label{con}
In the present work we studied classical bouncing behavior of the Universe in the framework of $f({\sf R},{\sf T})={\sf R}+h({\sf T})$ gravity theories. We assumed a single perfect fluid in a spatially flat, homogeneous and isotropic {\sf FLRW}  background. Having obtained the resulted field equations, we employed the concept of effective fluid (which is firstly introduced in~\cite{Shabani18}) via defining an effective energy density and pressure and also reformulating the field equations in terms of these fluid components. In this picture, one could recast the field equations of $f({\sf R},{\sf T})$ gravity for a real perfect fluid into {\sf GR} field equations for an effective fluid. It is also shown that in a modified gravity model the energy conditions are usually obtained by using the effective {\sf EMT}, not the one for real fluids. In $f({\sf R},{\sf T})$ gravity, the definitions for effective energy density and pressure have already been used to obtain the energy conditions~\cite{Sharif13}. The effective fluid has an {\sf EoS} of the form, $p_{(\sf{eff})}=\mathcal{Y}(\rho_{(\sf{eff})})$, which corresponds to an $h({\sf T})$ function. In this method one firstly specifies an effective {\sf EoS} or a condition on the effective components and then obtains the corresponding $h({\sf T})$ function and other cosmological quantities.
\par
It is also possible to make a link between $f({\sf R},{\sf T})$ gravity in effective picture and models which use some exotic or dark component with unusual {\sf EoS}. These models which have been widely discussed in the literature (to deal with some cosmological issues) are also called theories with \lq\lq{}generalized {\sf EoS}\rq\rq{}. The mathematical representation of effective components provides a setting within which unusual interactions of a real perfect fluid with gravitational field can be translated as the presence of an exotic fluid which admits the {\sf EoS} of the form $p_{(\sf{eff})}=\mathcal{Y}(\rho_{(\sf{eff})})$. In this paper we have shown that it is possible to recover generalized {\sf EoS} models which have been previously studied in the literature ( see e.g.,~\cite{Contreras17,Contreras16,Babichev05,Chavanis13,Bamba12}), in the framework of $f({\sf R},{\sf T})$ gravity. Therefore, the problem of exotic fluid in the context of generalized {\sf EoS} models which are mostly without a determined Lagrangian may be discussed in a Lagrangian based theory of gravity like $f({\sf R},{\sf T})$ gravity.

In the current research, we discussed four different bouncing models in $f({\sf R},{\sf T})$ gravity. We labeled them as the models {\sf A, B, C, D} and {\sf E} and briefly mentioned their main properties in Table~\ref{Tab1}. Each model can be specified either by an $h({\sf T})$ function or by an effective {\sf EoS}. Models {\sf A}-{\sf D} mimic an asymptotic de Sitter expansion in the far past and future of the bounce. The model {\sf A} corresponds to a constant effective pressure, $p_{(\sf{eff})}=\mathcal{P}$; for the model {\sf B} we have $p_{\sf{(eff)B}}=-\rho_{\sf{(eff)B}}/3+\sqrt{b_{\sf{B}}\rho_{\sf{(eff)B}}+d_{\sf{B}}}+e_{\sf{B}}$, the  model {\sf C} is specified by $p_{\sf{(eff)}C}=j_{\sf{C}} \rho_{(\sf{eff})C}+e_{\sf{C}}$, the model {\sf D} corresponds to  $p_{\sf{(eff)D}}=\rho_{\sf{(eff)D}}+\sqrt{b_{\sf{D}}\rho_{\sf{(eff)D}}+d_{\sf{D}}}+e_{\sf{D}}$ and finally the model {\sf E} obeys the {\sf EoS}, $p_{\sf{(eff)E}}=a_{\sf{E}}\rho_{\sf{(eff)E}}+\sqrt{b_{\sf{E}}\rho_{\sf{(eff)E}}+d_{\sf{E}}}+e_{\sf{E}}$, where the constants $b,~d$,~ $e$ and $j$ are written in terms of model parameters. In all models the matter density grows to a maximum value at the bounce which corresponds to a minimum for the scale factor. The effective density varies from zero at the bounce to a positive value in the far past and future of the bounce. The effective pressure varies between negative values; in model {\sf A}, it is a constant, in the model {\sf B} it increases at the bounce, in the models {\sf C} and {\sf E} it decreases and the model {\sf D} admits both behaviors. The effective {\sf EoS} has the property $-\infty<\mathcal{W}<-1$ when the bounce point is approached. The Hubble parameter satisfies $H(t)=0$ and $dH/dt>0$ at the event of bounce and also all its time derivatives have regular behavior for all models. Therefore, these bouncing solutions do not exhibit future singularities which are classified in the literature of cosmological solutions. We can consider the inherent exoticism hidden behind $f({\sf R},{\sf T})$ gravity in another way. As already we mentioned, this issue can be described as an unusual  interaction between gravitational field and normal matter or introducing an effective fluid. From the point of view of the energy conditions, in all discussed models the {\sf SEC} and {\sf NEC} are violated (note that for a normal fluid {\sf NEC} is not violated~\cite{Novello08}) near the bounce and the effective density gets minimized to zero. Such a result has been previously predicted in {\sf GR}~\cite{Molina99}. As discussed in~\cite{Novello08}, the exoticness can be understood as a minimization in the effective pressure. In the other words, a minimum in the effective energy density corresponds to a minimum in the scale factor. Such a behavior is permitted provided that $\mathcal{W}<-1$. Note that, for a normal matter, a minimum (maximum) compression leads to a minimum (maximum) energy density. Thus, in $f({\sf R},{\sf T})$ gravity an abnormal or effective fluid which leads to an uncommon balance in the density and pressure can be responsible for the bouncing behavior. An interesting feature of the bouncing solution in $f({\sf R},{\sf T})$ gravity is that one can construct solutions in which the {\sf SEC} is respected by the real perfect fluid. Such solutions cannot be found in {\sf GR}~\cite{Molina99}. Also note that the real perfect fluid with $w>-1$  never violates the {\sf NEC}. Therefore, we have solutions without the future singularities and all energy conditions can be respected by a real perfect fluid. By this discussion, one may use the definition of an (effective) phantom scalar field if one asks for the matter source to be reinterpreted as that of a scalar matter field. We obtained the equivalent scalar field $\phi_{\sf{(eff)}}(t)$ and its corresponding potential $V_{\sf{(eff)}}(t)$ in each case. Moreover, we have studied the dynamical system representation of these models. We found that the evolution of the Universe can be displayed by trajectories which initially start from an unstable state, passing through an unstable fixed point (the bounce event) and finally are absorbed by a stable point. The initial and final states are de-Sitter era in models {\sf A}, {\sf B}, {\sf C} and {\sf D} and the decelerated expanding Universe in model {\sf E}. Another important issue discussed in this work is related to the study of stability of bouncing solutions through scalar-type cosmological matter perturbations in the bouncing universe. Our numerical analysis of density perturbations for models {\sf A}, {\sf B} and {\sf C} revealed that, though a slight jump (depending on the initial conditions) at the bounce point, the amplitude of matter density perturbation ($\delta$) and perturbed potential ($\Phi$) behave regularly throughout the bounce phase. Therefore, since the time interval during which the fluctuations that occur within density contrast and perturbed field is short, the instabilities do not have enough time to grow to a significant magnitude. However, this case does not happen for the two remaining models.
\par
As the final remarks we should emphasize that our models were obtained by indicating different conditions on the effective density and pressure which led to different $h({\sf T})$ functions. This means that the models {\sf A, B, C, D} and {\sf E} are not the only possible models for the bouncing behavior. It is obvious that one can still choose other $h({\sf T})$ functions or consider other assumptions on the effective density and pressure to obtain new bouncing solutions (with even new features). Our aim was to show the existence of varieties of bouncing solutions in $f({\sf R},{\sf T})$ gravity and study their properties. Especially, our study was confined to the Lagrangians of type $f({\sf R},{\sf T})={\sf R}+h({\sf T})$ though other forms of Lagrangians can be investigated. The other issue is that our study was performed in the effective picture. In case such an approach is not taken seriously, one can think of it as only an alternative mathematical method. One can still investigate a nonsingular cosmological scenario without employing the equations which are written in terms of the effective quantities. In this case it is enough to assume a Lagrangian and solve the field equations to inspect for a bouncing solution. However, cosmological solutions for the $f({\sf R,T})$ gravity model presented here are not singularity free and as we observed under certain conditions, a class singular solutions could be obtained.

\end{document}